\documentclass[twocolumn]{jpsj3}
\usepackage{txfonts}
\usepackage{color}

\newcommand{\YbNiAl}{YbNi$_3$Al$_9$}
\newcommand{\YbNiCuAl}{Yb(Ni$_{1-x}$Cu$_x$)$_3$Al$_9$}
\newcommand{\CrNbS}{CrNb$_3$S$_6$}

\title{
Chiral Soliton Lattice Formation in Monoaxial Helimagnet Yb(Ni$_{1-x}$Cu$_x$)$_3$Al$_9$ 
}

\author{Takeshi Matsumura$^{1,2,6,7}$, Yosuke Kita$^{1}$, Koya Kubo$^{1}$, Yugo Yoshikawa$^{1}$, Shinji Michimura$^{3}$, Toshiya Inami$^{4}$, Yusuke Kousaka$^{5,6}$, Katsuya Inoue$^{5,6}$, and Shigeo Ohara$^{8}$}
\inst{$^{1}$Department of Quantum Matter, AdSM, Hiroshima University, Higashi-Hiroshima, Hiroshima 739-8530, Japan \\
$^{2}$Institute for Advanced Materials Research, Hiroshima University, Higashi-Hiroshima, Hiroshima 739-8530, Japan \\
$^{3}$Department of Physics, Faculty of Science, Saitama University, Saitama 338-8570, Japan\\
$^{4}$Synchrotron Radiation Research Center, National Institutes for Quantum and Radiological Science and Technology, Sayo, Hyogo 679-5148, Japan\\
$^{5}$Department of Chemistry, Graduate School of Science, Hiroshima University, Higashi-Hiroshima, Hiroshima 739-8526, Japan\\
$^{6}$Chirality Research Center (CResCent), Hiroshima University, Higashi-Hiroshima, Hiroshima 739-8526, Japan\\
$^{7}$Center for Emergent Condensed Matter Physics (ECMP), Hiroshima University, Higashi-Hiroshima, Hiroshima 739-8526, Japan\\
$^{8}$Department of Physical Science and Engineering, Graduate School of Engineering, Nagoya Institute of Technology, Nagoya 466-8555, Japan
}

\abst{
Helical magnetic structures and its responses to external magnetic fields in \YbNiCuAl, with a chiral crystal structure of the space group $R32$, have been investigated by resonant X-ray diffraction. 
It is shown that the crystal chirality is reflected to the helicity of the magnetic structure by a one to one relationship, indicating that there exists an antisymmetric exchange interaction mediated via the conduction electrons. 
When a magnetic field is applied perpendicular to the helical axis ($c$ axis), the second harmonic peak of $(0, 0, 2q)$ develops with increasing the field. 
The third harmonic peak of $(0, 0, 3q)$ has also been observed for the $x$=0.06 sample. 
This result provides a strong evidence for the formation of a chiral magnetic soliton lattice state, a periodic array of the chiral twist of spins, which has been suggested by the characteristic magnetization curve. 
The helical ordering of magnetic octupole moments, accompanying with the magnetic dipole order, has also been detected. 
}

\begin{document}
\maketitle

\section{Introduction}
Chirality is one of the most fundamental elements of symmetry in nature. 
It plays an important role in various phenomena ranging from biological functions to physical properties of inorganic substances.\cite{Wagniere07} 
In magnetic materials lacking the local inversion center for the two-ion exchange interaction, a spiral magnetic order is often stabilized due to the antisymmetric Dzyaloshinskii-Moriya (DM) interaction, giving rise to distinct physical properties.\cite{Dzyaloshinsky58,Moriya60} 
A simultaneous appearance of electric polarization with the spiral magnetic order is a typical manifestation of such effects.\cite{Katsura05} 
In chiral magnetic materials without both inversion and mirror symmetries, a helical magnetic order with a fixed sense of spin rotation can be stabilized. 
When a magnetic field is applied to such a system, a characteristic arrangement of topological spin structure is often stabilized through a competition between the Zeeman energy and the twisting force from the DM interaction. 
In cubic B20-type compounds such as MnSi with the space group $P2_13$, for example, the helical spin structure transforms into a hexagonal lattice condensate of magnetic skyrmions.\cite{Yu10a,Yu10b,McGrouther16}  
The skyrmion state is a long-ranged pattern of twisted spin arrangements realized in magnetic fields, which becomes more stable in two dimensional configurations in thin films.\cite{Seki12,Leonov16} 

In monoaxial chiral helimagnet such as CrNb$_3$S$_6$ (space group $P6_322$), when a magnetic field is applied perpendicular to the helical axis, the helical ground state transforms into a periodic array of incommensurate chiral spin twist, which separates the ferromagnetically aligned commensurate region.\cite{Kishine15,Togawa16,Togawa12,Togawa13,Togawa15} 
This is a nonlinear order of topological spin structure, and is called a chiral magnetic soliton lattice (CSL). 
These materials are expected to provide a new functionality which is operated by tuning the number of skyrmions or solitons in the sample.\cite{Tsuruta16,Wang17} 

\begin{figure}
\begin{center}
\includegraphics[width=8cm]{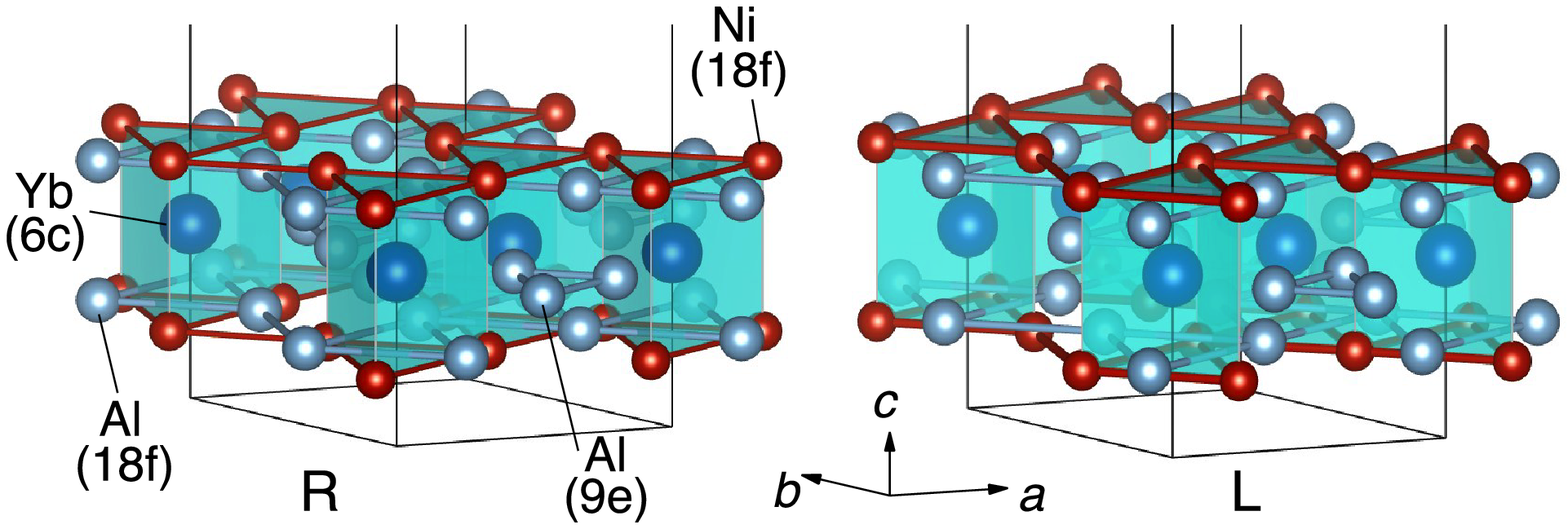}
\caption{(Color online) Crystal structure of \YbNiAl\ with the space group $R32$.\cite{Gladyshevskii93,Tobash11} 
We call the structure with $x$=0.3332, $y$=0.0056, $z$=0.08517 for the $18f$ site of Ni as right (R) and its mirror reflected structure as left (L) crystal. 
Only the Yb$_2$($6c$)+Al$_3$($9e$)+2Ni$_3$($18f$)+2Al$_3$($18f$) block layer at the bottom of the unit cell is shown. 
This block is shifted by $(2/3, 1/3, 1/3)$ and $(1/3, 2/3, 2/3)$.  
Other Al layers at the $6c$ and $9d$ sites are omitted. VESTA was used for drawing the figure.\cite{Momma11} 
}
\label{fig:cryst}
\end{center}
\end{figure}

In the present paper, we report on a new monoaxial chiral helimagnet system of \YbNiCuAl, in which the CSL state is expected to be realized. 
A rare earth compound \YbNiAl\ has a chiral crystal structure with the space group $R32$ (No. 155), which lacks both the space inversion and mirror symmetry.\cite{Gladyshevskii93,Tobash11} 
The main block of the crystal structure is picked up in Fig.~\ref{fig:cryst}.
Physical properties of \YbNiAl\ has been studied as a Yb-based heavy-fermion compound.\cite{Ohara11,Yamashita11,Yamashita12,Miyazaki12,Hirayama12,Utsumi12}
Yb ions form a two-dimensional honeycomb lattice in the $c$ plane, which is separated by $c/3$=9.121 \AA\ from the neighbouring Yb layer by five Al and two Ni layers. 
Since this is much larger than the nearest-neighbour distance of $a/\!\sqrt{3}$=4.199 \AA\ within a layer, the relation between the two dimensionality and the heavy fermion state has also been of interest. 
Detailed study of this compound from the viewpoint of chirality started from the discovery of a characteristic magnetization process which is reminiscent of a CSL state.\cite{Ohara14} 
By substituting Ni with Cu, they discovered that the $M(H)$ curve behaves like that of CrNb$_3$S$_6$,\cite{Miyadai83} in which the CSL state has certainly been identified.\cite{Togawa12,Togawa13,Togawa15} 

The exchange interaction in metallic \YbNiAl\ is considered to be of the Ruderman-Kittel-Kasuya-Yosida (RKKY) type. 
Since the crystal is chiral, there must be some antisymmetric contribution to the RKKY mechanism in the form of $\mib{D}_{ij}\cdot \mib{S}_i \times \mib{S}_j$, which is generally called the DM interaction.\cite{Dzyaloshinsky58,Moriya60} 
Since the microscopic mechanism given by Moriya is based on super-exchange interaction in insulators, 
the mechanism of RKKY-type DM interaction is an important subject to be studied. 
The question is how the crystal chirality is transferred to the spin system of $f$ electrons and conduction electrons. 
%This is another significance of the present study on this new compound \YbNiAl.  

\YbNiAl\ is a metallic compound which orders at $T_{\text{N}}$=3.5 K. 
It is reported that the magnetic structure of \YbNiAl\ is helical with a propagation vector $(0, 0, \sim\!\!0.8)$ and the moments lying in the $c$ plane. 
By applying a magnetic field perpendicular to the $c$ axis, the helimagnetic order jumps into a ferromagnetic state at a critical field of $H_{\text{c}}$=1 kOe.\cite{Miyazaki12,Yamashita12} 
By substituting Ni with Cu, both $T_{\text{N}}$ and $H_{\text{c}}$ increases. The anomalous $M(H)$ curve reminiscent of a CSL state is observed for $x$=0.06.\cite{Ohara14}  
However, the relationship between the crystal chirality and the magnetic helicity, Cu concentration dependence of the helimagnetic structure, and detailed dependences on temperature and magnetic field have not yet been studied. The aim of the present study is to clarify these properties experimentally by using resonant X-ray diffraction (RXD).

Experimentally, how to observe a chiral state has long been a subject of interest. 
Recent progress in the transmission electron microscope method has made it possible to capture real-space images of skyrmions and CSLs.\cite{Yu10a,Yu10b,McGrouther16,Seki12,Togawa12} 
It is also of fundamental importance to capture the spatially averaged structure as a Fourier transform of the real-space image by neutron and X-ray diffraction methods.
Polarized neutron diffraction is a powerful method to observe the helical magnetic structure and to determine the helicity.\cite{Ishida85,Yamasaki07} 
One drawback of this method is that it is difficult to perform the experiment in magnetic fields because the neutron spin state is affected by the magnetic field. 
X-ray diffraction, on the other hand, can be used both at zero field and in magnetic fields without any differences. 
Helicity of the magnetic spiral can be studied by using circularly polarized X-rays.\cite{Sutter97,Fabrizi09,Sagayama10} 
By utilizing resonance at an absorption edge of the magnetic element, the scattering cross section is enhanced, making it possible to detect signals from an ordered structure more efficiently.\cite{Hannon88} 
The magnetic skyrmion state in a chiral magnet has been detected by RXD.\cite{Yamasaki15} 
Furthermore, resonant scattering has a sensitivity to higher order anisotropy (multipole moments) of both magnetic and nonmagnetic nature.\cite{Hannon88,Lovesey05,Nagao06,Nagao10}
This sensitivity can sometimes be applied to determine crystal chirality by using circularly polarized beam.\cite{Kousaka09,Tanaka12a,Tanaka12b} 

%The first aim of the present work is to clarify the relationship between the crystal chirality and the magnetic helicity in \YbNiCuAl, which will provide direct evidence for the existence of DM interaction in this compound. The second aim is to detect signals from the CSL state in magnetic fields, which is proposed to exist by the magnetization measurement. 
This paper is organized as follows. 
In \S 2, the experimental procedure is described, including the details of the circularly polarized X-ray beam. 
The experimental results and the analyses are described in \S 3. 
First, in \S 3.1, the one to one relationship between the crystal chirality and the helimagnetic structure is described. 
Comparison of the experimental result with the helimagnetic structure model is performed in \S 3.2.  
In \S 3.3 and \S 3.4, the temperature and the magnetic field dependences of the helimagnetic order are presented. 
The resonant nature of the signal is described in \S 3.5. We show that the $E2$ resonance involves a signal from magnetic octupole, which  accompany with the helical order of the magnetic dipole. 
In \S 4, we discuss the origin of the octupole moment,  temperature dependent pitch of the helical structure, possibility of the CSL state in \YbNiCuAl. 
The present study will be summarized in \S 5.

\section{Experiment}
Single crystals of Yb(Ni$_{1-x}$Cu$_{x}$)$_3$Al$_9$ were prepared by an Al-flux method following the procedure as described in the literature.\cite{Ohara14} The starting Cu composition $x'$ sealed in a quartz ampoule was set five times the target composition $x$. The actual Cu concentration $x$ in the obtained crystal were checked by an electron-probe-microanalysis and were confirmed to follow the relation $x\sim 0.2 x'$, as reported previously.\cite{Ohara14} 
We also checked the sample quality by the magnetic susceptibility, magnetization, and electrical resistivity measurements, and obtained consistent results as those reported in the literatures.\cite{Ohara14,Yamashita12} 

\begin{figure}
\begin{center}
\includegraphics[width=8cm]{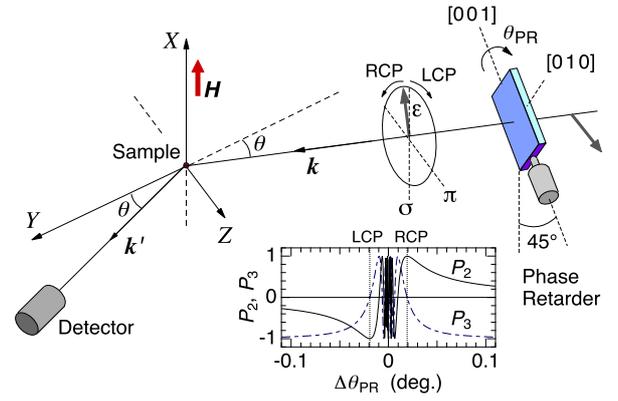}
\caption{(Color online) Scattering configuration of the experiment with a phase retarder system inserted in the incident beam. 
The inset figure shows the $\Delta\theta_{\text{PR}} = \theta_{\text{PR}} -  \theta_{\text{B}}$ dependence of the polarization state using the Stokes parameters $P_2$ and $P_3$. The vertical dotted lines represent the positions of LCP and RCP states. 
The beam is depolarized in the region around $\Delta\theta_{\text{PR}}\approx 0$. }
\label{fig:Config}
\end{center}
\end{figure}

RXD experiments were performed at BL22XU at SPring-8.  
The $c$-plane surfaces of the samples were mirror polished, and the samples were mounted in a vertical-field 8 Tesla superconducting cryomagnet equipped with a $^3$He cryostat insert, so that the $c$ axis was perpendicular to the magnetic field and coincided with the scattering vector $\mib{k}' - \mib{k} \parallel \hat{Z}$.  
The scattering geometry is shown in Fig.~\ref{fig:Config}. 
Incident X-ray energy was tuned at the $L_3$ edge of Yb. 

We used a diamond phase retarder system to tune the horizontally polarized incident beam to a circularly polarized state.\cite{Inami13} 
By rotating the angle of the diamond phase plate, $\theta_{\text{PR}}$, about the 220 Bragg angle $\theta_{\text{B}}$, 
where the scattering plane is tilted by $45^{\circ}$, a phase difference arises between the $\sigma$ and $\pi$ components of the transmitted beam. 
The phase difference is approximately proportional to $1/(\theta_{\text{PR}} - \theta_{\text{B}})$. 
This allows us to tune the incident linear polarization to right-handed circular polarization (RCP) and left-handed circular polarization (LCP) by changing $\Delta\theta_{\text{PR}} = \theta_{\text{PR}} - \theta_{\text{B}}$. 
The polarization state of the incident beam as a function of $\Delta\theta_{\text{PR}}$ is shown in Fig.~\ref{fig:Config} using the Stokes parameters $P_2$ ($+1$ for RCP and $-1$ for LCP) and $P_3$ ($+1$ for $\sigma$ and $-1$ for $\pi$ linear polarization).\cite{Lovesey96} 
In the horizontal scattering plane configuration in the present experiment, the incident linear polarization is $\pi$ when $\Delta\theta_{\text{PR}}$ is large. 
$P_1$ ($+1$ for $45^{\circ}$ and $-1$ for $-45^{\circ}$ linear polarization) is zero in the present setup. 

We define RCP as $\varepsilon_{\text{R}} = (\varepsilon_{\sigma} + i \varepsilon_{\pi} ) e^{i(\mib{k}\cdot \mib{r} - \omega t)}$ 
and LCP as $\varepsilon_{\text{L}} = (\varepsilon_{\sigma}- i \varepsilon_{\pi}) e^{i(\mib{k}\cdot \mib{r} - \omega t)}$.
We have checked the helicity of the incident photon after transmitting the phase retarder 
%, i.e., the relation of $P_2>0$ when $\Delta\theta_{\text{PR}}>0$, 
by measuring the resonant scattering intensity of a forbidden reflection from a $P6_122$-type CsCuCl$_3$, where the intensity ratio between RCP and LCP X-rays depends on the reflection index and is exactly determined by the $P6_122$ space group.\cite{Kousaka09}

Crystal chirality of the sample was determined by using a laboratory based X-ray diffraction system (Bruker APEX-II) and it was also checked in the RXD experiment at the beam line. 
In the former method, using a Mo $K_{\alpha}$ X-ray beam, the Flack parameter was deduced by analyzing the intensities of many reflections, which resulted in either 0 (R) or 1 (L). 
Our definition of the crystal chirality is shown in Fig.~\ref{fig:cryst}. 
Then, R and L samples were selected for each Cu concentration for the RXD experiment. 
At the beam line, energy dependences of the $(1, 1, 24)$ and $(\bar{1}, \bar{1}, 24)$ fundamental Bragg-peak intensities were measured around the absorption edge of Yb, which is shown in Fig.~\ref{fig:Edep1124} in the Appendix B.
The spectrum exhibited a contrasting energy dependence at the edge depending on the chirality of the crystal. 
It was consistent with the calculated spectrum assuming the predetermined crystal chirality, 
confirming that the irradiated spot of the sample in the RXD experiment has exactly the same chirality as the one determined in the laboratory X-ray diffraction.  
It was also confirmed at the beam line that the $(1, 0, 3n+1)$ reflections are allowed and the $(1, 0, 3n)$ and $(1, 0, 3n-1)$ reflections are forbidden, which is shown in Fig.~\ref{fig:theta10L}. 
This fact guarantees the three-fold symmetry of the sample about the $c$ axis. If the forbidden reflection were observed, it means that the $[1\,1\,0]$ axis is mixed with the $a$ axis due to the stacking fault by $60^{\circ}$.

\section{Results and Analysis}
\subsection{Crystal chirality and the helical magnetic structure}
\begin{figure}
\begin{center}
\includegraphics[width=8cm]{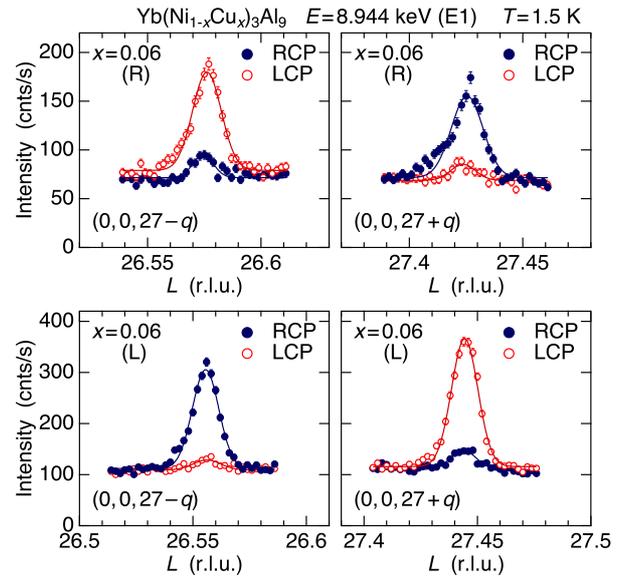}
\caption{(Color online) Reciprocal space scan along $(0, 0, L)$ around the $(0, 0, 27)$ fundamental peak for the $x$=0.06 sample with right and left chirality using RCP and LCP X-rays. Solid lines are the fits with Gaussian functions. }
\label{fig:LscansPR}
\end{center}
\end{figure}

First, we demonstrate that the crystal chirality and the helicity of the helical magnetic structure has a one to one relationship. 
A typical example is shown in Fig.~\ref{fig:LscansPR} for the R and L crystals of $x$=0.06. 
The reciprocal scans along (0, 0, $L$) around the $(0, 0, 27)$ fundamental Bragg peak using RCP and LCP X-rays exhibit opposite behavior for the R and L crystals and for the $(0, 0, 27-q)$ and $(0, 0, 27+q)$ peaks. 
In the R(L) crystal at $(0, 0, 27-q)$, the intensity is strong(weak) for LCP and weak(strong) for RCP. This relation is reversed at $(0, 0, 27+q)$. 
The X-ray energy of 8.944 keV corresponds to the $E1$ ($2p\leftrightarrow 5d$) resonance. The energy dependence of the intensity demonstrating the resonance feature will be shown later. 
In Fig.~\ref{fig:LscansPR}, the peaks are observed at an incommensurate wave vector of $q$=0.445 for the L crystal and $q$=0.425 for the R crystal. 
This difference in the $q$ value is due to the difference in the sample quality, i.e., a subtle difference in the Cu concentration, and has nothing to do with the chirality.

More detailed information can be extracted from the $\Delta\theta_{\text{PR}}$ scans, which are shown in Fig.~\ref{fig:PRscansE1}. 
This figure shows not only the intensity relations for the RCP and LCP X-rays at $(0, 0, 27\pm q)$ but also the whole feature of the incident polarization dependence as a function of $\Delta\theta_{\text{PR}}$. The incident polarization state varies with $\Delta\theta_{\text{PR}}$ as shown in the inset of Fig.~\ref{fig:Config}. 
We again observe from this measurement that the crystal chirality and the helicity of the magnetic structure has a one to one relationship. 
In addition, the relationship does not change with the Cu concentration between $x$=0 and $x$=0.06. 
The solid lines in the figures represent the calculated curve expected from the helical magnetic structure with the moments lying in the $c$ plane and propagating along the $c$ axis. 
Next, we describe the analysis of the above experimental results.  

\subsection{Magnetic structure}
There are two Yb atoms for the $6c$ site of the $R32$ space group: Yb-1 at $\mib{d}_1 = (0, 0, z)$ and Yb-2 at $\mib{d}_2 = (0, 0, \bar{z})$, 
where $z$=0.167$ \sim 1/6$. In the present single-$\mib{q}$ magnetic structure, the magnetic moment $\mib{\mu}_{1,j}$ and $\mib{\mu}_{2,j}$ of Yb-1 and Yb-2, 
respectively, on the $j$th lattice point at $\mib{r}_j=(n_1, n_2, n_3)$, $(n_1+2/3, n_2+1/3, n_3+1/3)$, and $(n_1+1/3, n_2+2/3, n_3+2/3)$, where $n_1$, $n_2$, and $n_3$ are integers, are generally expressed as
\begin{subequations}
\begin{align}
\mib{\mu}_{1,j} &= \mib{m}_1 e^{i\mib{q}\cdot\mib{r}_j} + \mib{m}_1^* e^{-i\mib{q}\cdot\mib{r}_j} \,, \\
\mib{\mu}_{2,j} &= \mib{m}_2 e^{i\mib{q}\cdot\mib{r}_j} + \mib{m}_2^* e^{-i\mib{q}\cdot\mib{r}_j} \,, 
\end{align}
\end{subequations}
where $\mib{m}_1$ and $\mib{m}_2$ are the magnetic amplitude vectors of Yb-1 and Yb-2, respectively. 
In the present case of Yb(Ni$_{1-x}$Cu$_{x}$)$_3$Al$_9$, since the moments are expected to be ordered within the $c$ plane, $\mib{m}_1$ and $\mib{m}_2$ can generally be written as $\mib{m}_1 = m_1 (\hat{\mib{x}} + e^{i\varphi} \hat{\mib{y}})$ and $\mib{m}_2 = m_2 e^{i\delta} (\hat{\mib{x}} + e^{i\varphi} \hat{\mib{y}})$, 
where $\hat{\mib{x}}$ and $\hat{\mib{y}}$ represent the unit vectors along the $x$ and $y$ axis, which are taken perpendicular to the $c$ ($z$) axis. 
The irreducible representation of the $\mib{m}$-vector for $\mib{q}=(0, 0, q)$, where $q$ is an incommensurate value, is written by $\hat{\mib{x}} \pm i \hat{\mib{y}}$. 
Therefore, $\varphi$ is either $\pi/2$ or $-\pi/2$.  
The phase difference between $\mib{\mu}_{1,j}$ and $\mib{\mu}_{2,j}$ is represented by $\delta$. 
The above expression of the magnetic structure can be reduced to 
\begin{subequations}
\begin{align}
\mib{\mu}_{1,j} &= m_1 \bigl\{ \hat{\mib{x}} \cos \mib{q}\cdot\mib{r}_j + \hat{\mib{y}} \cos (\mib{q}\cdot\mib{r}_j + \varphi) \bigr\} \,, \\
\mib{\mu}_{2,j} &= m_2 \bigl\{ \hat{\mib{x}} \cos (\mib{q}\cdot\mib{r}_j + \delta) + \hat{\mib{y}} \cos (\mib{q}\cdot\mib{r}_j + \varphi + \delta) \bigr\} \,.
\end{align}
\end{subequations}
Since $\varphi=\pm \pi/2$, the above expression describes a perfect helical structure with a helicity $\pm 1$ in which the adjacent Yb-1 (or Yb-2) moments on the neighboring layers along the $c$ axis make a fixed angle of $2\pi q/3$. 
%When $\delta=-q/3$, the Yb moments in a layer are ferromagnetically coupled. 
On the other hand, the angle between the Yb-1 and Yb-2 moments within a layer, the former at $(0, 0, z)$ and the latter at $(2/3, 1/3,1/3)+(0, 0, \bar{z})$ in Fig.~\ref{fig:cryst}, is described by the parameter $\delta$.
It is reported to be 20.5$^{\circ}$ for $x$=0.\cite{MunakataICM} 
Note that our experimental results presented in this paper, which were collected only along the $(0, 0, L)$ line, are not sensitive to determine the $\delta$ value. 
%At the same time, the following data analysis is not affected by the $\delta$ value.

\begin{table}
\caption{Relation among the phase $\varphi$, Fourier component $\mib{m}$, scattering vector $\mib{Q}$, magnetic structure factor $\mib{Z}_{\text{dip}}^{(1)}$, sign of $C_2$, and the crystal chirality. }
\label{tbl:1}
\begin{tabular}{cccccc}
\hline
$\varphi$ & $\mib{m}$ & $\mib{Q}$ & $\mib{Z}_{\text{dip}}^{(1)}$ & $C_2$ & Crsytal\\
\hline
$\pi/2$ & $\hat{\mib{x}}+ i\hat{\mib{y}}$ & $(0, 0, 3n \pm q)$ & $(1, \pm i, 0)$ & $ \pm $ & R\\
%            &   & $(0, 0, 3n-q)$ & $(1, -i, 0)$ & $ - $  & R\\
$-\pi/2$ & $\hat{\mib{x}} - i\hat{\mib{y}}$ & $(0, 0, 3n \pm q)$ & $(1,  \mp i, 0)$ & $ \mp $ & L \\
%            &   & $(0, 0, 3n-q)$ & $(1, i, 0)$ & $ + $ & L\\
\hline
\end{tabular}
\end{table}

\begin{figure}
\begin{center}
\includegraphics[width=8cm]{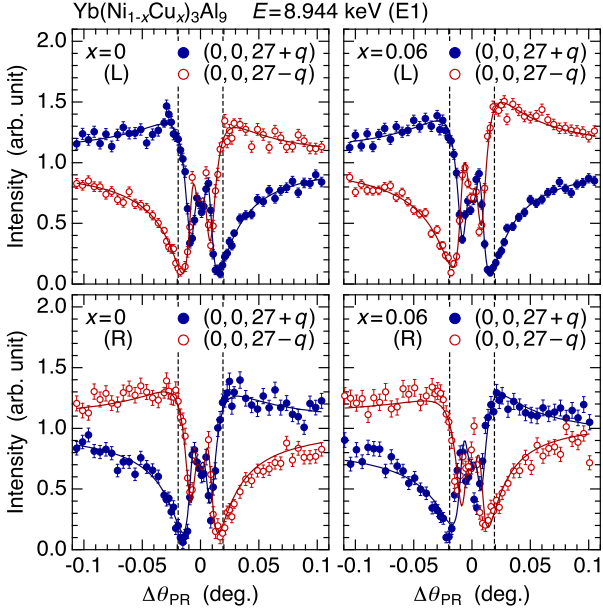}
\caption{(Color online) Incident polarization ($\Delta\theta_{\text{PR}}$) dependence of the intensities at $(0,0,27\pm q)$ for $x$=0 and $x$=0.06 crystals with right and left handed chirality. Background has been subtracted. 
Solid lines are the calculations described in the text.}
\label{fig:PRscansE1}
\end{center}
\end{figure}

\begin{figure}
\begin{center}
\includegraphics[width=7cm]{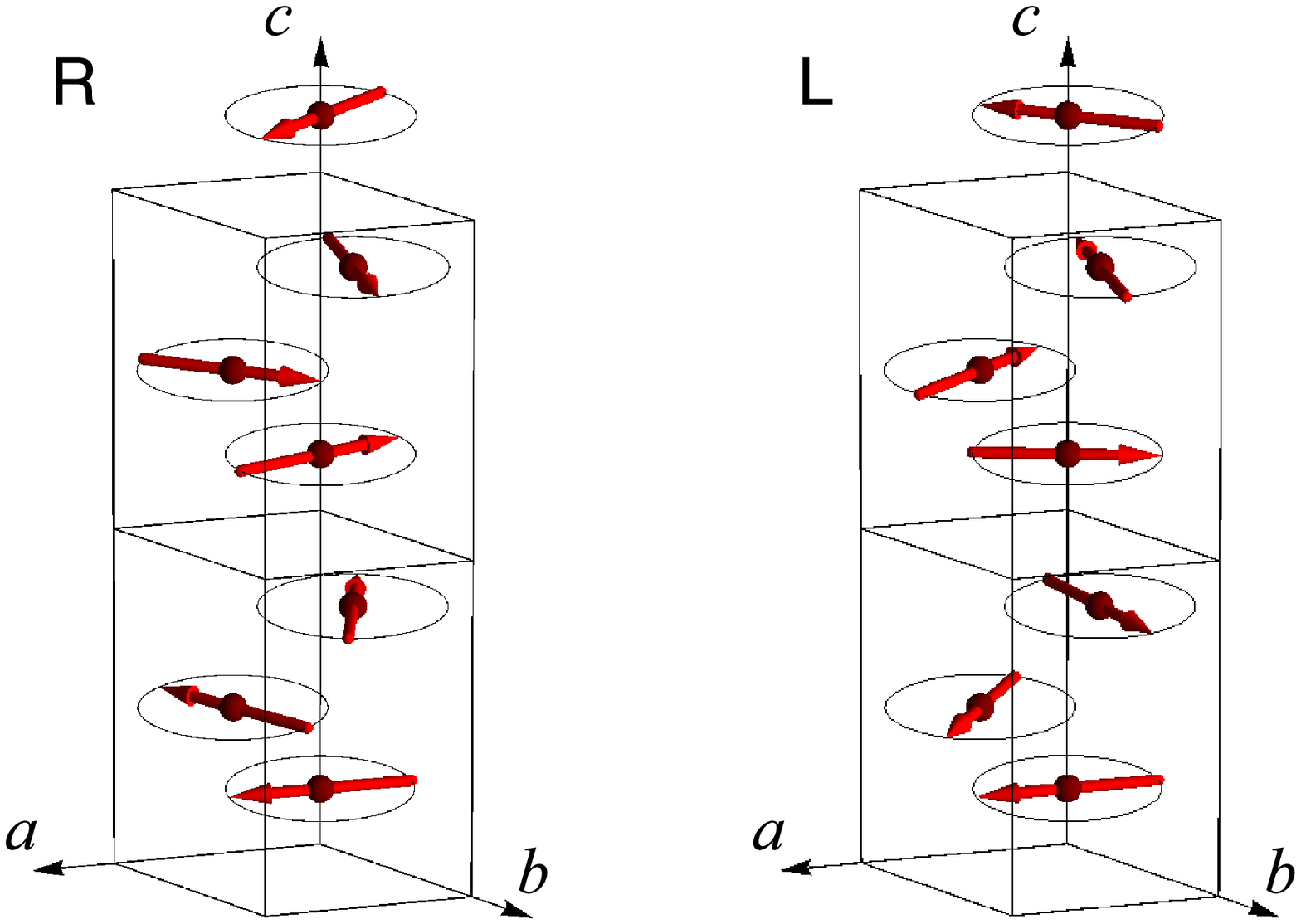}
\caption{(Color online) Magnetic structure of \YbNiCuAl\ for $x$=0.06. Only the moments on the Yb-1 sites are shown. 
%R crystal: $\varphi=\pi/2$, $\mib{m}_{q}=(\hat{x} + i\hat{y})$, $F_{m, \pm q}=(1, \pm i, 0)$. 
%L crystal: $\varphi=-\pi/2$, $\mib{m}_{q}=(\hat{x} - i\hat{y})$, $F_{m, \pm q}=(1, \mp i,0)$.
} 
\label{fig:Magst}
\end{center}
\end{figure}

The $E1$ resonant scattering amplitude from magnetic dipole order is proportional to  
$i (\mib{\varepsilon}' \times \mib{\varepsilon})\cdot \mib{Z}_{\text{dip}}^{(1)}$,\cite{Hannon88,Lovesey05} 
where 
\begin{equation}
\mib{Z}_{\text{dip}}^{(1)} = \sum_{j,d} \mib{\mu}_{j,d} e^{-i\mib{Q}\cdot(\mib{r}_{j}+\mib{d})}
\end{equation}
represents the magnetic structure factor at $\mib{Q}=\mib{k}' - \mib{k}$. 
At $\mib{Q}=(0, 0, 3n \pm q)$, $\mib{Z}_{\text{dip}}^{(1)}=(1, \pm i, 0)$ when $\varphi=\pi/2$, 
and $\mib{Z}_{\text{dip}}^{(1)}=(1, \mp i, 0)$ when $\varphi=-\pi/2$. 
The scattering amplitude matrix, as defined in the Appendix A, for the magnetic structure factor $(1, \pm i, 0)$ is expressed as
\begin{equation}
\hat{F}_{E1}=\begin{pmatrix}
0 & i \cos \theta \\  -i \cos\theta &  \mp  \sin 2\theta
\end{pmatrix} \,.
\end{equation}
Then, following the method described in the Appendix A, we can calculate the coefficients $C_n$ ($n=0 \sim 3$) defined in Eq.~(\ref{eq:CrossSec2}), which expresses the scattering cross section.  
The parameters $C_2/C_0$ and $C_3/C_0$, as normalized by the total cross section $C_0$, correspond to the intensity term proportional to $P_2$ and $P_3$, respectively.  
When $C_2>0$, the intensity is stronger for the RCP ($\Delta\theta_{\text{PR}}>0$). 
For $\mib{Q}=(0, 0, 27 \pm q)$, when $\varphi=\pi/2$, $C_2/C_0=\pm 0.71$ and $C_3/C_0=-0.47$ are obtained. 
The observed intensity is a superposition of the $P_2$ and $P_3$ terms. 
The experimental data in Fig.~\ref{fig:PRscansE1} show that, at $\mib{Q}=(0, 0, 27 + q)$, $C_2 > 0$ in the R-crystal and $C_2 < 0$ in the L-crystal. 
This means that $\varphi=\pi/2$ ($\mib{m}=\hat{\mib{x}}+ i\hat{\mib{y}}$) in the R-crystal 
and $\varphi=-\pi/2$ ($\mib{m}=\hat{\mib{x}}- i\hat{\mib{y}}$) in the L-crystal. 
These relations are summarized in Table.~\ref{tbl:1}. 
The magnetic structure is shown in Fig.~\ref{fig:Magst} for $x$=0.06. 
The Yb-2 moment is not shown because the relative angle with the Yb-1 moment is unknown.
The solid lines in Fig.~\ref{fig:PRscansE1} are the calculations using the $C_2/C_0$ and $C_3/C_0$ values calculated above, which agree well with the experimental data.

\subsection{Temperature dependence}
\begin{figure}
\begin{center}
\includegraphics[width=8cm]{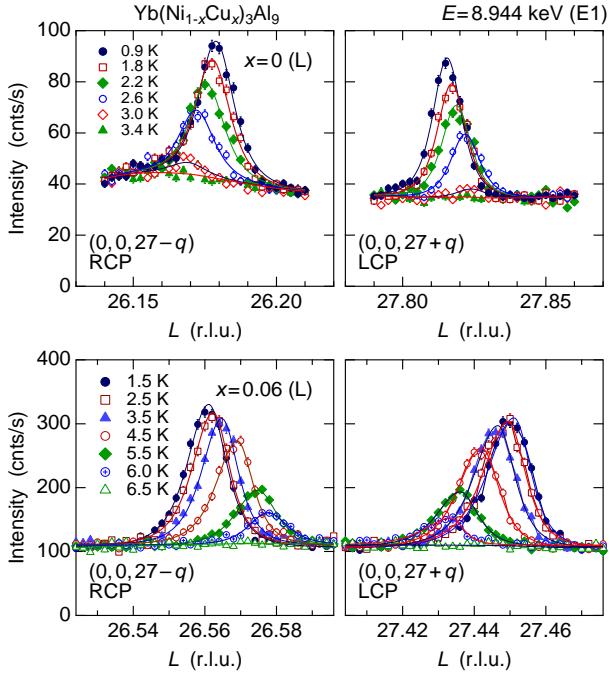}
\caption{(Color online) Temperature dependence of the peak profile for $x$=0 (L) and $x$=0.06 (L). }
\label{fig:TdepLscan}
\end{center}
\end{figure}

Figure \ref{fig:TdepLscan} shows the $(0, 0, L)$ peak profile for $x$=0 (L) and $x$=0.06 (L) samples. 
With increasing $T$, the intensity decreases and vanishes at $T_{\text{N}}$, indicating that the resonant signal is of magnetic origin. 
In addition, the peak position shifts with the temperature. 
It is also noteworthy that the direction of the peak shift for $x$=0.06 is opposite to that for $x$=0. 
It depends on the Cu concentration $x$, but does not depend on the chirality of the crystal. 

\begin{figure}
\begin{center}
\includegraphics[width=8.5cm]{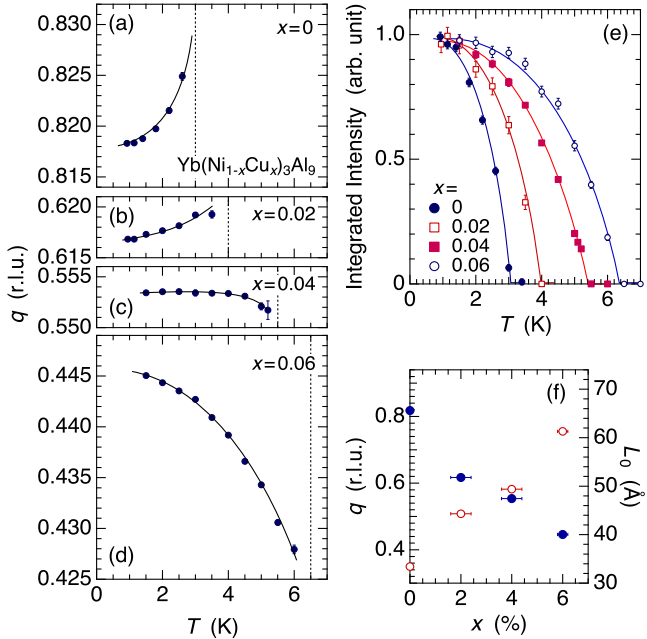}
\caption{(Color online) (a)--(d) : Temperature dependence of the $q$ value of the propagation vector. The vertical dotted line represents $T_{\text{N}}$.
(e): Temperature dependence of the normalized intensity. (f): Cu concentration dependence of the $q$ value (closed circles) and the pitch $L_0=c/q$ (open circles) at the lowest temperature. }
\label{fig:TdepParams}
\end{center}
\end{figure}

The parameters obtained from the $(0, 0, L)$ scans are summarized in Fig.~\ref{fig:TdepParams}. 
For $x$=0, the $q$ value decreases with decreasing $T$ and saturates at $\sim 0.818$ at the lowest temperature, 
indicating that the helical structure is incommensurate with the lattice. 
The $T$-dependence of the $q$ value becomes weak for $x$=0.02. 
Surprisingly, at $x$=0.04, the direction of the $T$-dependence is reversed and the $q$ value increases with decreasing $T$. 
At $x$=0.06, the $T$-dependence becomes strong again. 
It seems that the direction of the shift in the $q$ value is reversed at around $x$=0.03, where $q\sim 0.6$. 

Figure \ref{fig:TdepParams}(e) shows the $T$ dependence of the integrated intensity for all the Cu concentration $x$. 
$T_{\text{N}}$ increases roughly proportional to $x$, which is consistent with the literature.\cite{Ohara14,Oharapc} 
In addition, the shift of the $q$ value mentioned above, regardless of its direction, is almost proportional to the $T$-dependence of the intensity, 
which reflects the development of the ordered moment. 
This suggests that the $q$-shift is associated with the magnitude of the ordered moment. 
In Fig.\ref{fig:TdepParams}(f), we show the $x$ dependence of the $q$ value and the pitch $L_0=c/q$ at the lowest temperature. 
$L_0$ should be compared with the inter-layer distance of $c/3\sim 9.1$ \AA. 
The angle between the moments on neighboring layers is calculated by $2\pi q/3$, i.e., $98.2^{\circ}$, $74.0^{\circ}$, $66.4^{\circ}$, and $53.4^{\circ}$, for $x$=0, 0.02, 0.04, and 0.06, respectively, at the lowest temperature. 
%The $x$ value represents the nominal one. There can be an error for the 2\% and 4\% sample as indicated by the error bar. It roughly seems that the $q$ value changes linearly with $x$. 

\subsection{Magnetic field dependence}
\begin{figure}
\begin{center}
\includegraphics[width=8cm]{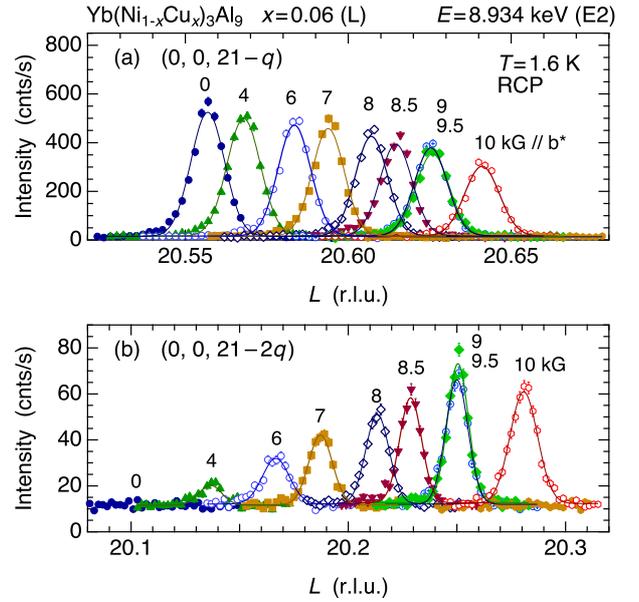}
\caption{(Color online) Magnetic field dependence of the peak profile for $x$=0.06 measured with the RCP photon at the $E2$ resonance energy. 
(a) First harmonic peak at $(0, 0, 21 - q)$. (b) Second harmonic peak at $(0, 0, 21 - 2q)$.  Field is applied perpendicular to the $c$ axis ($\parallel b^*$). }
\label{fig:LscanH006}
\end{center}
\end{figure}

Figure \ref{fig:LscanH006}(a) shows the magnetic field dependence of the peak profile of $(0, 0, 21 - q)$ for the $x$=0.06 (L) sample, measured at the $E2$  ($2p\leftrightarrow 4f$) resonance energy of 8.934 keV, where the signal to noise ratio was much higher than that at the $E1$ resonance. The energy dependence of the resonant signal will be shown later. 
We used the RCP photon with stronger scattering intensity for the L-crystal than the LCP photon. 
The result for the second harmonic peak measured at $(0, 0, 21 - 2q)$ is shown in Fig. \ref{fig:LscanH006}(b).
With increasing $H$, the peak position shifts to the fundamental Bragg peak at $(0, 0, 21)$ and the intensity of the first harmonic ($q$) peak gradually decreases. 
On the other hand, the intensity of the second harmonic ($2q$) peak, which does not exist at zero field, gradually increases with increasing $H$. 

\begin{figure}
\begin{center}
\includegraphics[width=8cm]{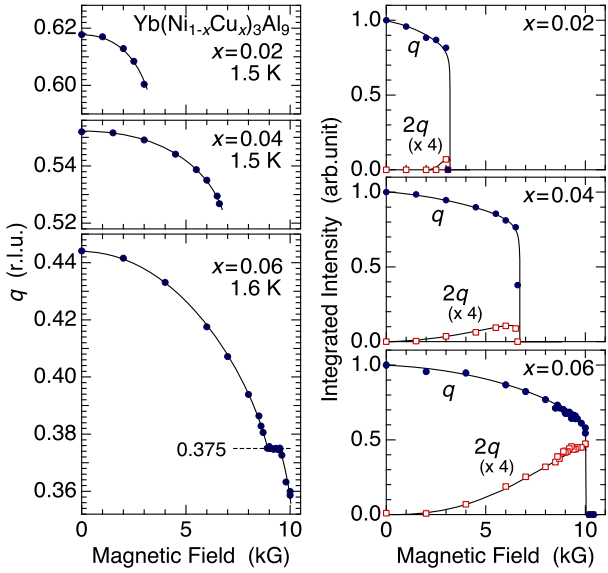}
\caption{(Color online) Magnetic field dependence of the $q$ values (left column) and the integrated intensities (right column). The integrated intensities for the $2q$ peak is multiplied by four. }
\label{fig:HdepParams}
\end{center}
\end{figure}

Magnetic field dependence of the $q$ values and the integrated intensities of the $q$ and $2q$ peaks for $x$=0.02, 0.04, and 0.06, are summarized in Fig.~\ref{fig:HdepParams}. 
Although all these measurements have been performed using the L-crystals, the results do not depend on the crystal chirality. 
For all the Cu concentrations, it is commonly observed that the $q$ value decreases with increasing $H$, which becomes more rapid at higher fields on approaching the critical field $H_{\text{c}}$. 
Note that $q$ does not decrease continuously to zero, but jumps to zero at $H_{\text{c}}$. 
The integrated intensity of the $q$ peak gradually decreases with increasing $H$ and also jumps to zero at $H_{\text{c}}$. 
These results show that the transition at $H_{\text{c}}$ is of first order. 
It is also a common characteristic that the $2q$ peak is gradually induced with increasing $H$. 
It is observed even for the $x$=0.02 sample with $\mu_0 H_{\text{c}}$=3 kG. 

Another noteworthy result is the locking-in behavior of the helical magnetic propagation vector at $q$=0.375=3/8 between 9 to 9.5 kG for $x$=0.06. 
As observed in Fig.~\ref{fig:LscanH006}, the peak position does not change in this field region. 
The $2q$ peak is also stuck at $2q$=3/4. 
This result shows that there exists some coupling between the helimagnetic structure and the lattice, although the magnetic anisotropy in the $c$ plane is considered to be very small. 
%At the present stage, however, it is not clear which component, $q$ or $2q$, is responsible for the locking-in behavior. 
There is no hysteresis in this behavior, and can be observed both in field increasing and decreasing processes. 
It is also interesting that the lock-in does not seem to exist at $q$=3/7=0.4286 at around 5 kG. 
This suggests that the lock-in is more associated with the $2q$ peak, which develops at high fields, rather than the $q$ peak existing from zero field to high fields. 

\begin{figure}
\begin{center}
\includegraphics[width=8cm]{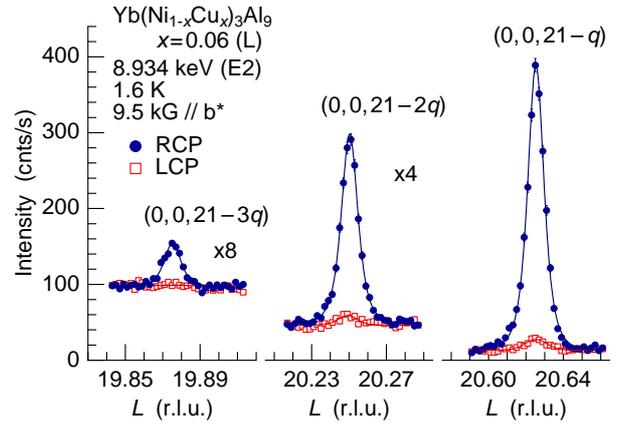}
\caption{(Color online) $L$-scan profiles for the higher harmonic peaks of $(0, 0, 21-nq)$ $(n=1,2,3)$ observed at 9.5 kG for $x$=0.06 (L). }
\label{fig:Harmonics}
\end{center}
\end{figure}

At 9.5 kG just before the critical field of 10 kG to the ferromagnetic state in the $x$=0.06 (L) sample, 
we searched for more higher order harmonics and successfully detected the third harmonic peak. 
The result is shown in Fig.~\ref{fig:Harmonics}. 
For all the harmonics, the scattering intensity for the RCP photon is stronger than the LCP photon, 
indicating that the magnetic helicity of the modulated structure giving the higher harmonic is the same as that of the original helical structure of the first harmonic. 
The fourth harmonic peak was too weak to be recognized above the background. 
The integrated intensity of the $2q$ and $3q$ peaks are approximately 6 and 50 times weaker, respectively, than that of the $q$ peak. 
%The integrated intensity of the $n$th harmonic follows an approximate line proportional to $1/6^n$. 

The $H$-dependence of the $q$ value directly shows that the pitch of the helimagnetic structure increases with increasing $H$. 
In addition, the appearance of the higher harmonic peaks shows that some additional structure is superimposed with increasing $H$. 
It is remarkable that the $H$-dependence of the $q$ and $2q$ peak intensities shown in Fig.~\ref{fig:HdepParams} is very similar to the calculation for the chiral sine-Gordon model (see Fig. 25 of Ref.~\citen{Kishine15} or Fig. 13 of Ref.~\citen{Togawa16}). 
This result, as well as the $H$-dependence of $q$ which is associated with $L_0/L_{\text{CSL}}$ (Fig. 13 of Ref.~\citen{Kishine15} or Fig. 38 of Ref.~\citen{Togawa16}), 
strongly suggests that a CSL state is formed in \YbNiCuAl, especially for $x$=0.06.

\subsection{Energy spectrum}
\begin{figure}
\begin{center}
\includegraphics[width=8.5cm]{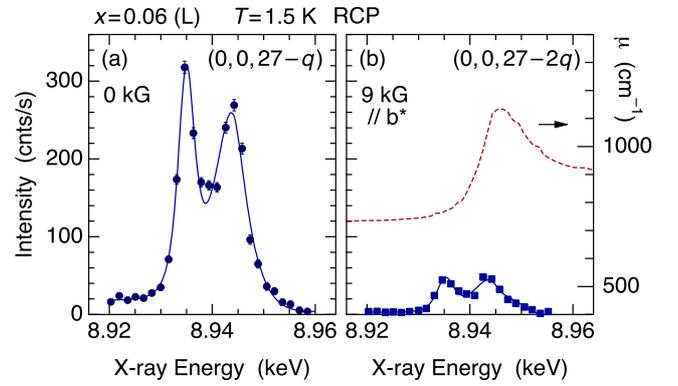}
\caption{(Color online) Energy dependence of (a) the $(0, 0, 27-q)$ peak-intensity at zero field, and (b) the $(0, 0, 27-2q)$ peak-intensity at 9 kG for $x$=0.06 (L) measured with RCP incident photon after background subtraction and absorption correction. 
Absorption coefficient obtained from the fluorescence spectrum is shown in the right panel.  
}
\label{fig:Edepx006}
\end{center}
\end{figure}

Figure \ref{fig:Edepx006} shows the energy dependence of the $(0, 0, 27-q)$ peak at 0 kG and the $(0, 0, 27-2q)$ peak at 9 kG measured for $x$=0.06 (L). 
Two resonant peaks are well separated at 8.934 keV and 8.944 keV. The former can be assigned to the $E2$ resonance ($2p \leftrightarrow 4f$) and the latter to the $E1$ resonance ($2p \leftrightarrow 5d$) peak. 
In the tail of the lower energy side of the $(0, 0, 27-q)$ peak we can see a weak nonresonant magnetic scattering. 
The energy dependence for the second harmonic peak also exhibit the resonances at the $E2$ and $E1$ energies. 
No nonresonant signal has been observed. This shows that the resonant signal of the second harmonic can be ascribed to the helical magnetic order itself, and not to a possible lattice deformation induced by the magnetic order. 
These features of the energy dependence are commonly observed for other Cu concentrations.

\subsection{Contribution of magnetic octupole to the $E2$ resonance}
\begin{figure}
\begin{center}
\includegraphics[width=8cm]{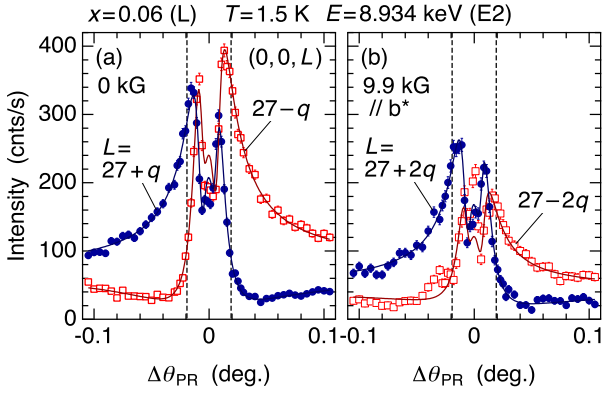}
\caption{(Color online) $\Delta\theta_{\text{PR}}$ dependence of the $E2$ resonance intensity for (a) $(0, 0, 27\pm q)$ at 0 kG and (b) $(0, 0, 27\pm 2q)$ at 9.9 kG. 
The vertical dashed lines represent the positions of LCP and RCP states. Solid lines are the fits using Eq.~(\ref{eq:CrossSec2}). }
\label{fig:PRscanE2}
\end{center}
\end{figure}

\begin{figure}
\begin{center}
\includegraphics[width=8cm]{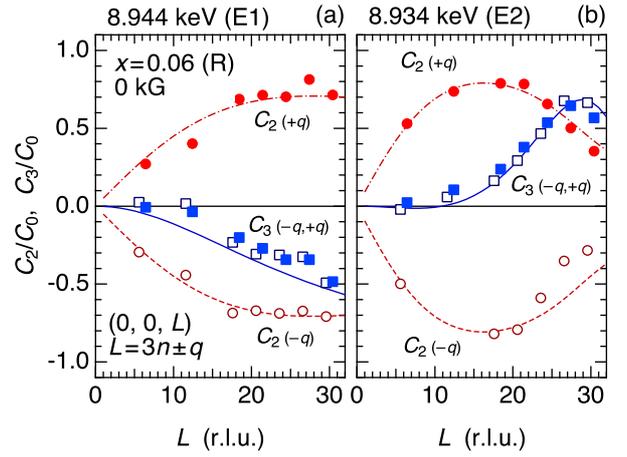}
\caption{(Color online) $L$ ($=3n \pm q$) dependence of the parameters $C_2/C_0$ and $C_3/C_0$ at zero field obtained from the $\Delta\theta_{\text{PR}}$ scans for the $x$=0.06 (R) sample at (a) $E1$ resonance and (b) $E2$ resonance. 
Solid, dashed, and dot-dashed curves represent the calculations 
of $C_3/C_0$ ($L=3n \pm q$), $C_2/C_0$ ($L=3n - q$), and $C_2/C_0$ ($L=3n + q$), respectively, for (a) $E1$ resonant scattering from helimagnetic dipole order and (b) $E2$ resonant scattering from helimagnetic dipole and octupole order.}
\label{fig:LdepC2C3}
\end{center}
\end{figure}

Figure \ref{fig:PRscanE2} shows the $\Delta\theta_{\text{PR}}$ dependence of the $(0, 0, 27-q)$ peak-intensity at zero field and the $(0, 0, 27-2q)$ peak-intensity at 9.9 kG at the $E2$ resonance energy. 
We notice that the $\Delta\theta_{\text{PR}}$ dependence is apparently different from that of Fig.~\ref{fig:PRscansE1} for the $E1$ resonance. 
The $E2$ intensity is strongly enhanced when $\Delta\theta_{\text{PR}}$ is tuned to the LCP or RCP positions. 
The result that the two $\Delta\theta_{\text{PR}}$ dependences are very similar shows that the structure factors for $q$ and $2q$ peaks are almost the same. 
By fitting these $\Delta\theta_{\text{PR}}$ dependences using Eq.~(\ref{eq:CrossSec2}), we obtain three parameters of $C_0$, $C_2$ and $C_3$. 
The parameters obtained for various $\mib{Q}$ vectors for the first harmonic peak from $(0, 0, 6\pm q)$ to $(0, 0, 30\pm q)$ for the $x$=0.06 (R) sample is shown in Fig.~\ref{fig:LdepC2C3}. 
The parameters $C_2$ and $C_3$ are normalized by $C_0$. 
Note that the sign of $C_2$ in Fig.~\ref{fig:LdepC2C3} is opposite to the one deduced in Fig.~\ref{fig:PRscanE2} because the crystal chirality is opposite. 
With respect to the $E1$ resonance, the $Q$-dependence data in Fig.~\ref{fig:LdepC2C3}(a) can be well reproduced by the scattering amplitude of 
magnetic dipole, 
$i (\mib{\varepsilon}' \times \mib{\varepsilon})\cdot \mib{Z}_{\text{dip}}^{(1)}$, as explained in \S 3.2.
The calculated $Q$-dependences of the parameters are shown by the lines in Fig.~\ref{fig:LdepC2C3}(a). 

The problem is that the results of $E2$ resonance in Fig.~\ref{fig:PRscanE2}(a) and Fig.~\ref{fig:LdepC2C3}(b) cannot be reproduced by the scattering from the magnetic dipole moment.
Using Eq.~(\ref{eq:XfacE2dip}), the $E2$ scattering-amplitude matrix for the magnetic dipole with the structure factor $(1, \pm i, 0)$ is expressed as 
\begin{equation}
\hat{F}_{E2}=\begin{pmatrix}
\mp \sin 2\theta & i \cos 3\theta \\  -i \cos 3\theta &  \mp  2 \sin 4\theta
\end{pmatrix}\,.
\end{equation}
For $\mib{Q}=(0, 0, 27 \pm q)$, when $\varphi=\pi/2$ for the R-crystal in the present case, $C_2/C_0=\mp 0.79$ and $C_3/C_0=0.51$ are obtained by calculation. 
However, the sign of $C_2$ is opposite to the experimental result. 
In addition, since the $\cos 3\theta$ in $\hat{F}_{\pi\sigma'}$ and $\hat{F}_{\sigma\pi'}$ changes sign at $\theta=\pi/6$, $C_2$ in Fig.~\ref{fig:LdepC2C3}(b) should change sign around $L$=20, which clearly disagrees with the experimental result.\cite{S1} 
Therefore, we cannot explain the $E2$ resonance by the scattering from magnetic dipole only. 

The $Q$-dependence of $C_2$ and $C_3$ for $E2$ can be reproduced by considering the scattering from magnetic octupole. 
We use the following model. 
First, the structure factor of magnetic dipole (rank-1) for $\mib{Q}=(0, 0, 3n \pm q)$ for the R-crystal is represented by 
\begin{align}
\mib{Z}_{\text{dip}}^{(1)} &= (1, \pm i, 0) \,. \label{eq:ZT1u-1}
\end{align}
Second, with respect to the magnetic octupole (rank-3), we take into account the $T^{\alpha}_{x,y,z}$ ($T_{1u}$) and $T^{\beta}_{x,y,z}$ ($T_{2u}$) moments. 
%Both of them have $x$, $y$, and $z$ components. 
The structure factors of these moments, which are compatible with the helical magnetic dipole order, should be written as  
\begin{subequations}
\begin{align}
\mib{Z}_{\alpha}^{(3)} &= \sqrt{3} A e^{i\phi} (1, \pm i , 0) \,, \label{eq:ZT1u-3}\\
\mib{Z}_{\beta}^{(3)} &= \sqrt{5} A e^{i\phi} (1, \mp i , 0) \,,\label{eq:ZT2u-3}
\end{align}
\end{subequations}
where $A$ and $\phi$ represent the relative amplitude and phase, respectively, of the octupolar structure factor with respect to those of $\mib{Z}_{\text{dip}}^{(1)}$. 
By putting $A=0.46$ and $\phi=-0.79 \pi$, the data are well explained as shown by the calculated curves in Fig.~\ref{fig:LdepC2C3}(b). 
The $\Delta\theta_{\text{PR}}$ dependence in Fig.~\ref{fig:PRscanE2}(a) can also be reproduced by using these parameters. 

The detailed relationship of the coefficients in Eqs.~(\ref{eq:ZT1u-3}) and (\ref{eq:ZT2u-3}) comes from the irreducible representation of the order parameter in the $D_3$ point group.\cite{S2} 
Although we use the cubic coordinate for the intensity calculation, we should analyze the order parameter using the hexagonal coordinate.
The in-plane dipole moments, $J_x$ and $J_y$, for example, belong to the two dimensional $E_u$ representation in the $D_3$ point group. 
The octupole moments with the $E_u$ representation, on the other hand, are expressed as linear combinations of $T^{\alpha}$ and $T^{\beta}$.  
They are written as $T^{\gamma}_x=(-\sqrt{3}T^{\alpha}_x - \sqrt{5}T^{\beta}_x)$ and $T^{\gamma}_y=(-\sqrt{3}T^{\alpha}_y +\sqrt{5}T^{\beta}_y)$.  
Therefore, the ordered structure of the octupole moments given by Eqs.~(\ref{eq:ZT1u-3}) and (\ref{eq:ZT2u-3}) is equivalent to the helical octupole ordering expressed by 
$T^{\gamma}_x + i T^{\gamma}_y$, which has the same symmetry as the helical magnetic dipole order expressed by $J_x + i J_y$. 
Although the opposite sign of $\mib{Z}_{\beta, y}^{(3)}$ in Eq.~(\ref{eq:ZT2u-3}) is rather a tricky result, it is not surprising. 

Other representations are given by $T^{\delta}_x=(\sqrt{5}T^{\alpha}_x - \sqrt{3}T^{\beta}_x)$ ($A_{2u}$) and $T^{\delta}_y=(\sqrt{5}T^{\alpha}_y +\sqrt{3}T^{\beta}_y)$  ($A_{1u}$). 
$T_{xyz}$ and $T^{\beta}_z$ also constitute another $E_u$ representation in the $D_3$ group. 
Although these moments generally need to be included, the data were successfully explained without taking into account these moments.

\section{Discussions}
\subsection{Helical order of octupole moments}
The ground multiplet of $J=7/2$ for Yb$^{3+}$ splits into four Kramers doublets in the crystalline electric field of \YbNiAl, represented by a point group $D_3$. 
Since the magnetic moments prefer to lie in the $c$ plane, the crystal field ground state is likely to be composed mainly of $|\frac{7}{2}, \pm \frac{1}{2} \rangle$, which is isotropic in the $c$ plane. 
In this two-dimensional space of $|\frac{7}{2}, \pm \frac{1}{2} \rangle$, the $x$-component matrices of $\hat{J}_x$ ($E_{u}$ dipole) and $\hat{T}^{\gamma}_{x}$ [$x(5z^2-r^2)$ type, $E_{u}$ octupole] are expressed in the same form as  
\begin{equation}
\hat{J}_x = 2 \begin{pmatrix} 0 & 1 \\ 1 & 0 \end{pmatrix}, \;\;\;\; 
\hat{T}^{\gamma}_x = -15\sqrt{\frac{3}{2}} \begin{pmatrix} 0 & 1 \\ 1 & 0 \end{pmatrix}.
\end{equation}
Therefore, when a magnetic dipole moment $\langle J_x \rangle$ is induced, the $\langle T^{\gamma}_x \rangle$ octupole arises simultaneously. 
When a helical magnetic dipole order of $\langle J_x + iJ_y \rangle$ occurs, the helical magnetic octupole order of $\langle T^{\gamma}_x + iT^{\gamma}_y \rangle$ with the same helicity also arise. 
The matrix elements of the $T_{xyz}$ and $T^{\beta}_{z}$ octupoles ($E_{u}$), $T^{\delta}_x$ ($A_{2u}$), and $T^{\delta}_y$ ($A_{1u}$) all vanish in the $|\frac{7}{2}, \pm \frac{1}{2} \rangle$ space. This is the reason we did not need to include these moments in the analysis in \S 3.6. 
Also, there is no degree of freedom for the quadrupole (rank-2) moments, which guarantees the reasoning that the orbital contribution is not included in the second harmonic signal.

\begin{figure}
\begin{center}
\includegraphics[width=8cm]{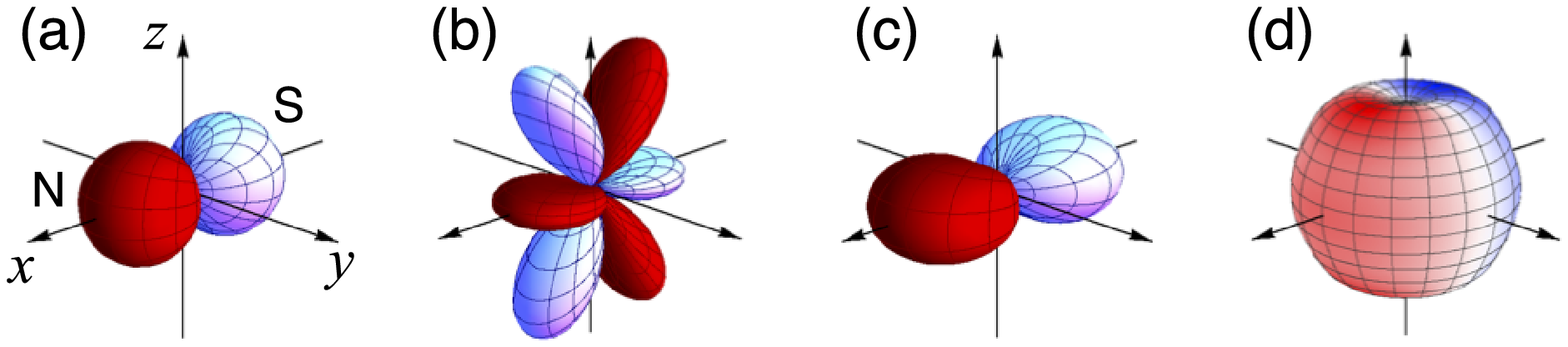}
\caption{(Color online) Magnetic charge distribution representing (a) a magnetic dipole $J_x$, (b) a magnetic octupole $-T^{\gamma}_x$, (c) a summation of $J_x$ and $-T^{\gamma}_x$. (d) charge and magnetic charge density of the $|7/2, \pm 1/2\rangle$ state of Yb$^{3+}$, possessing a magnetic dipole moment $\langle J_{x} \rangle$.\cite{Kusunose08} }
\label{fig:Octupole}
\end{center}
\end{figure}

The $T^{\gamma}$-type magnetic octupole reflects an aspherical charge distribution of the $|\frac{7}{2}, \pm \frac{1}{2} \rangle$ states. 
%If the charge distribution were spherical, the magnetized state can be simply expressed by an arrow as shown in Fig.~\ref{fig:Magst}, representing the magnetic dipole moment. 
%This is equivalently represented by a magnetic charge distribution shown in Fig.~\ref{fig:Octupole}(a).  
%However, this simple picture is not applicable to the $|\frac{7}{2}, \pm \frac{1}{2} \rangle$ case for Yb$^{3+}$, where the orbital is more expanded in the $c$ ($xy$) plane. 
%In such a case, there also arise an aspherical magnetic moment density, which is represented by magnetic octupole moments. 
Figure~\ref{fig:Octupole} shows schematics of the magnetic charge distributions. 
The superposition of the $J_x$-dipole and the the $T^{\gamma}$-octupole, shown in Figs.~\ref{fig:Octupole}(a) and \ref{fig:Octupole}(b), respectively, results in the magnetic charge distribution shown in Fig.~\ref{fig:Octupole}(c). 
%Note that the sign of $\langle T^{\gamma}_x \rangle$ is opposite to that of $\langle J_x \rangle$. 
We see less magnetic moment density along the $c$ ($z$) axis, which is associated with the less charge density of the $|\frac{7}{2}, \pm \frac{1}{2} \rangle$ state along the $c$ axis as shown in Fig.~\ref{fig:Octupole}(d). 

At the present stage, it is not clear whether the octupole moments affect any physical property in \YbNiCuAl\ or not. 
It should be remarked, however, that the strength of the inter-ionic multipolar interaction is rank independent in the RKKY mechanism.\cite{Shiina97,Shiba99} 
As a result, the higher rank multipoles play important roles in the physical property and can equally be a primary order parameter as is typically realized in the cubic Ce$_x$La$_{1-x}$B$_6$ system, in which the dipole, quadrupole, and octupole moments are active as independent degrees of freedom.\cite{Matsumura12,Matsumura14}
In \YbNiCuAl, by contrast, the active moments in the doublet ground state are $(J_x, J_y)$ or $(T^{\gamma}_x, T^{\gamma}_y)$, which both belong to the $E_u$ representation and are not independent. 
The issue is how much the physical property is affected by the octupolar interaction in comparison with the dipolar interaction.

\subsection{Temperature dependent helical magnetic structure}
As shown in Fig.~\ref{fig:TdepParams}, the propagation vector changes with temperature in proportion to the magnitude of the ordered moment. 
This suggests that the RKKY exchange interaction itself changes with the evolution of the ordered moment.\cite{JM91} 
The magnetic propagation vector generally reflects the $\mib{q}$ vector where the exchange interaction $J(\mib{q})$, the Fourier transform of $J_{ij}$, takes the maximum. 
$J(\mib{q})$ for the RKKY interaction is associated with the local $c$-$f$ exchange interaction and $\chi(\mib{q})$ for the conduction electron system. 
Unfortunately, little is known yet about the Fermi surface of \YbNiCuAl. 
When a helimagnetic order develops on the Yb sites, there arises a perturbation of exchange field to the conduction electron system, which is also described by the same $\mib{q}$ vector. 
%$\mathcal{H}_{cf}=-J \sum_{i} \delta(\mib{r} - \mib{R}_i) \mib{s} \cdot \mib{S}_i$, 
%$\mathcal{H}_{ff}\propto J \sum_{\mib{q}} \chi(\mib{q}) e^{i \mib{q}\cdot (\mib{R}_i - \mib{R}_j)} \mib{S}_i \cdot \mib{S}_j$
As a result, a gap appears in the region of the Fermi surface where $\varepsilon_{\mib{k}'}=\varepsilon_{\mib{k}+\mib{q}}$ is satisfied.\cite{Watson68} 
This gap slightly modifies $\chi(\mib{q})$ and $J(\mib{q})$, resulting in a shift of the $\mib{q}$ vector.\cite{Elliot64} 
The $\mib{q}$-shift from the original value of $\mib{q}_0$ just below $T_{\text{N}}$ becomes almost proportional to the development of the ordered moment. 
A similar $T$-dependence of the $\mib{q}$ vector has also been reported in GdSi, GdNi$_2$B$_2$C and GdPd$_2$Al$_3$.\cite{Feng13a,Feng13b,Detlefs96,Inami09}.
In any case, concerning the most noteworthy phenomenon that the direction of the $\mib{q}$-shift changes below and above $x$=0.03, 
we have no explanation yet. 
Knowledge on the band structure and the Fermi surface is required.

If there were magnetic anisotropy, the $T$-dependence of the $\mib{q}$ vector would be more complicated as observed in rare earth metals.\cite{JM91} 
An anisotropy in the $c$ plane would cause a squaring up and the appearance of the third harmonic peak. 
This effect also causes the $\mib{q}$ vector to shift from $\mib{q}_0$. 
However, in such a case, the $\mib{q}$-shift becomes proportional to $(T_{\text{N}} - T)^2$, which is different from the present case in \YbNiCuAl.\cite{JM91,Sato94} 
Furthermore, the $|7/2,\pm 1/2\rangle$ states in \YbNiCuAl\ are almost isotropic in the $c$ plane.
%, and we may neglect the magnetic anisotropy. 

\subsection{Formation of chiral soliton lattice}
The results at zero field (Figs.~\ref{fig:LscansPR} and \ref{fig:PRscansE1}) show that the helicity of the helimagnetic structure in \YbNiCuAl\ has a one to one relation with the crystal chirality. 
This shows that there indeed exists a DM exchange interaction, lifting the degeneracy between the left- and right-handed helical magnetic structures. 
The situation is similar to the well investigated compound \CrNbS, and \YbNiCuAl\ can also be recognized as a monoaxial chiral helimagnet. 
The shift of the $\mib{q}$-vector and the appearance of the higher harmonic peaks in magnetic fields (Figs.~\ref{fig:LscanH006}, \ref{fig:HdepParams}, and \ref{fig:Harmonics}) suggest the formation of a CSL, a characteristic outcome of a monoaxial chiral helimagnet. 

Theoretical study on the CSL state has been performed using the chiral sine-Gordon model, which has been successfully applied to \CrNbS.\cite{Kishine15}
In \CrNbS, the application of the sine-Gordon model, where continuous variables are introduced to treat the spin arrangement, is justified because 
the modulation length of the helimagnetic structure $L_0$ ($\sim 480$ \AA) is much larger than the inter-layer distance $c_0$ =12.1 \AA. 
On the other hand, in \YbNiCuAl, it should be remarked that the longest $L_0$ of 61 \AA\ in $x$=0.06 is only 6.7 times larger than $c_0$=9.1 \AA\ (=$c/3$), which questions the simple application of the continuous model. However, without an appropriate theory at the present stage, it is worth comparing the results with the sine-Gordon model. 

\begin{figure}
\begin{center}
\includegraphics[width=8cm]{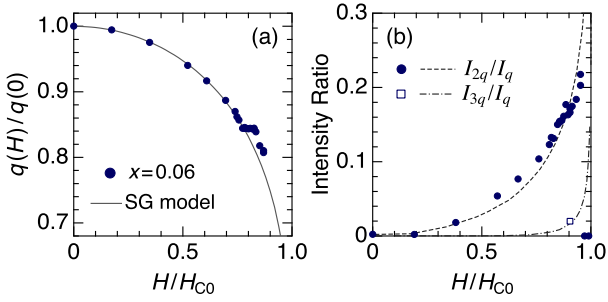}
\caption{(Color online) (a) Comparison of the experimental data of $q(H)/q(0)$ for $x=0.06$ with the $L_0/L_{\text{CSL}}$ curve obtained from the chiral sine-Gordon (SG) model.\cite{Kishine15,Togawa16} 
$\mu_0 H_{C0}$ is set to be 11.5 kG. 
(b) Comparison of the experimentally obtained intensity ratios of $I_{2q}/I_{q}$ (filled circles) and $I_{3q}/I_{q}$ (open square) with the the calculated ratio for the sine-Gordon model, using $\mu_0 H_{C0}$=10.5 kG. }
\label{fig:SGmodel}
\end{center}
\end{figure}

\begin{figure}
\begin{center}
\includegraphics[width=8cm]{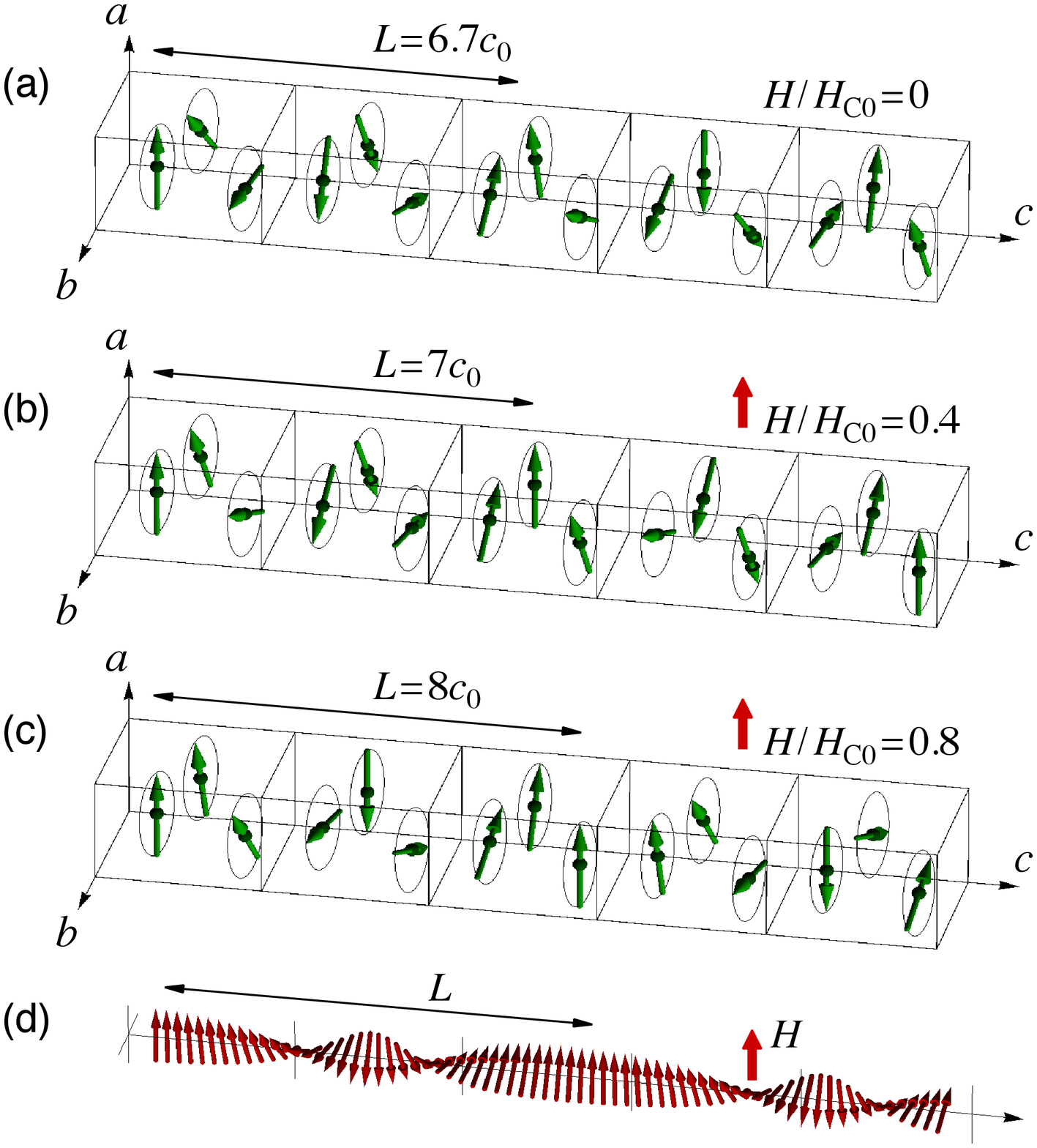}
\caption{(Color online) 
(a)--(c) Magnetic structures of \YbNiCuAl\ for $x$=0.06 (L) in magnetic fields expected from the chiral sine-Gordon model. 
Only the moments on Yb-1 sites are shown. 
(a) Helimagnetic ordered state at zero field with $q$=0.445 and the helical pitch of $L$=$6.7c_0$. The same as Fig. \ref{fig:Magst}. 
(b) CSL state for $H/H_{C0}$=0.4, corresponding to $q$=3/7 and $L$=$7c_0$ at around 5 kG. 
(c) CSL state for $H/H_{C0}$=0.8, corresponding to $q$=3/8 and $L$=$8c_0$ between 9.0 and 9.5 kG. 
(d) CSL state for continuous one-dimensional spin model corresponding to (c), $H/H_{C0}$=0.8. 
%(d) Schematic of the CSL state for continuous one-dimensional spin model corresponding to (c), $H/H_{C0}$=0.8. The virtual spins in the intermediate region between the Yb layers are represented by a different color. 
}
\label{fig:magstHSG}
\end{center}
\end{figure}

In Fig.~\ref{fig:SGmodel}(a), we compare the $H$-dependence of the normalized propagation vector $q(H)/q(0)$ for $x=0.06$ with the calculation of $L_0/L_{\text{CSL}}$ for the sine-Gordon model.\cite{Kishine15,Togawa16} 
We introduced a critical field of $\mu_0 H_{C0}$=11.5 kG so that the data fit to the calculation. 
The data points well follow the theoretical curve up to $H/H_{C0}\sim 0.85$, where the first order transition to the ferromagnetic state takes place. 
In Fig.~\ref{fig:SGmodel}(b), we compare the intensity ratios of $I_{2q}/I_{q}$ and $I_{3q}/I_{q}$ with the theoretical curve by using a slightly different $\mu_0 H_{C0}$ of 10.5 kG. 
These intensity ratios also follow the theoretical curve well. 
Although the calculation is for neutron diffraction, the result can almost equally be applied to the intensity of RXD. 
These agreements between experiment and theory strongly suggest that the CSL state is actually formed in the \YbNiCuAl\ system.

It is a problem, however, that the first order transition at $H/H_{C0}\sim 0.85$ is not explained by the theory, 
probably because the sine-Gordon model is too simple to describe the actual magnetic exchange interactions in the \YbNiCuAl\ system. It may be associated with the short period of the original helimagnetic structure. 
It is interesting to compare with the long period system of CrNb$_3$S$_6$, in which the CSL state survives against the ferromagnetic state up to $H/H_{C0}\sim 0.97$.\cite{Togawa12} 
Also, the lock-in phenomenon is not explained because the coupling with the lattice, other than the basal plane anisotropy, is not included in the theory. 
At the same time, we need to keep in mind why the continuous model of the sine-Gordon theory looks so applicable below $H/H_{C0}\sim 0.85$ to the short period system of \YbNiCuAl. This is an important problem to be studied in future. 

Putting aside the applicability of the theory, we show in Fig.~\ref{fig:magstHSG}(a)--(c) the theoretically expected magnetic structures in magnetic fields for $x$=0.06 (L). 
Of special interest is the locked-in structure with $q$=3/8 between 9 and 9.5 kG, which is shown in Fig.~\ref{fig:magstHSG}(c). 
Another structure with $q$=3/7=0.4286, which is expected to appear at around 5 kG, is shown in Fig.~\ref{fig:magstHSG}(b) for comparison. 
Since it may be difficult to catch the CSL structure from these figures because of the discontinuity in the spin arrangement, we show for reference in Fig.~\ref{fig:magstHSG}(d) the CSL state for a continuous one-dimensional spin model corresponding to the CSL state of Fig.~\ref{fig:magstHSG}(c). 
The CSL structure of Fig.~\ref{fig:magstHSG}(c) gives rise to all the harmonics of $q$, $2q$, $3q$, $\cdots$. 
It also reproduces the $\Delta\theta_{\text{PR}}$-scan data for the $2q$ peak shown in Fig.~\ref{fig:PRscanE2}(b) because the structure factor is written in the same way as $(1, \pm i, 0)$.  
These consistencies with the data suggest that the CSL structure of Fig.~\ref{fig:magstHSG}(c) is actually realized. 
If this is the case, there should be no reason that the $q$=3/7 structure of Fig.~\ref{fig:magstHSG}(b) is not stabilized. 
Experimentally, however, the $q$=3/7 structure does not seem to be locked in at around 5 kG as observed in Fig.~\ref{fig:HdepParams}. 
This shows that the lock-in behavior is strongly associated with the CSL formation. 
The CSL with $q$=3/8 at 9 kG locks in the lattice because the CSL is well developed, whereas the CSL with $q$=3/7 is not locked in because it is still weak at 5 kG.

\subsection{Magnetic exchange and DM interactions}
There are many problems concerning the magnetic exchange and DM interactions in this system. 
The sine-Gordon theory is based on the following one-dimensional model for a layered monoaxial chiral helimagnet with ferromagnetic nearest-neighbor exchange and DM interaction, 
\begin{equation}
\mathcal{H}=-J_1 \sum_{i} \mib{S}_i \cdot \mib{S}_{i+1} -\mib{D}\cdot\sum_{i} \mib{S}_i \times \mib{S}_{i+1} - \sum_{i} \mib{S}_i \cdot \mib{H}\,,
\end{equation}
where $\mib{S}_i$ lies in the $xy$ plane, $\mib{D}$ is along the $z$ axis, and $\mib{H}$ is applied in the $xy$ plane. 
At $H=0$, a helimagnetic order with $q_0 = c_0^{-1} \arctan (D/J_1)$ is formed, where $c_0$ is the interlayer distance. 
If we simply apply this model to  \YbNiCuAl, 
to obtain $q_0$ values of 0.42, 0.54, and 0.63 for $x$=0.06, 0.04, and 0.02, respectively, $D/J_1$ requires to be 1.2, 2.1, and 3.9, respectively, which are unrealistically large. 
For $x$=0, in addition, it is impossible to reproduce $q_0$=0.82 from this model because the angle between spins on neighboring layers is larger than 90$^{\circ}$. 
To explain these large $q_0$ values, it is necessary to include more long-ranged exchange interactions mediated by the conduction electrons. 
It is more reasonable to consider that $q_0$ is determined mainly by $J_1$ $(>0)$, $J_2$ $(<0)$, and further interactions, 
with which $J(q)$ takes a maximum at $q_0$. 
If we consider $J_1$ and $J_2$, for example, $J(q)=J_1 \cos q c_0 + J_2 \cos 2 q c_0$, and $q_0$ is given by $c_0^{-1} \arccos (-J_1/4J_2)$. 
The degeneracy of $\pm q_0$ is lifted by the DM term. 

However, there still remains a problem. 
If $q_0$ is determined by the maximum of $J(q)$, it is expected that $T_{\text{N}} \propto J(q_0)$ and $H_{\text{C}} \propto J(q_0)-J(0)$. 
When $q_0$ approaches 0 (ferromagnetic state) with increasing the Cu concentration, 
$J(q_0)-J(0)$ should also become small because $J(q)$ is a symmetric function of $q$.\cite{S3} 
However, while $T_{\text{N}}$=6.5 K for $x$=0.06 is twice as large as that for $x=$0, $\mu_0 H_{\text{C}}$=10 kG for $x$=0.06 is ten times as large as that for $x$=0. 
These facts are not explained by the simplistic model above and show that the situation behind the helimagnetic order of \YbNiCuAl\ is not so simple. 
Firstly, the interaction between Yb-1 and Yb-2 moments should be considered. 
More importantly, the fact that $H_{\text{C}}$ significantly increases with increasing $x$, i.e., with decreasing $q_0$ and $J(q_0)-J(0)$, 
suggests that the nonlinear effect of the CSL formation in magnetic fields plays an important role in keeping the helimagnetic structure and preventing the system from the transition to the ferromagnetic state. 
A possibility of the DM interaction to increase with increasing the Cu concentration should also be considered.

\section{Summary}
We performed resonant X-ray diffraction experiments to investigate the helical magnetic structure of \YbNiCuAl\ with the space group $R32$, which lacks both inversion and mirror symmetries and allows existence of left and right handed crystal structure. 
\begin{itemize}
\item
First, we showed that the magnetic structure is described by an incommensurate propagation vector $(0, 0, q)$, where $q$=0.818, 0.617, 0.553, and 0.445 for $x$=0, 0.02, 0.04, and 0.06, respectively, at the lowest temperature. 
The magnetic moments lie in the $c$ plane. 

\item
Second, we clarified that the helicity of the magnetic structure has a one to one relation with the crystal chirality, indicating that there exists an antisymmetric exchange interaction mediated by the conduction electrons, i.e., the RKKY mechanism. 
A theoretical study is required to provide further insight into the physical mechanism.  

\item
Third, when a magnetic field is applied perpendicular to the helical axis, the $q$ value decreases (the helical pitch increases) and, simultaneously, the higher harmonic peaks of $2q$ and $3q$ are induced with increasing the field. 
This behavior, especially for the $x$=0.06 case, coincides with the calculation for a chiral sine-Gordon model, which describes the formation of the chiral soliton lattice (CSL) state in a monoaxial helimagnet. This coincidence provides a strong evidence for the CSL formation in \YbNiCuAl, which has been suggested by the magnetization measurement. 
However, since the helical pitch is not much longer than the inter-layer spacing, it is questionable if the spin arrangement could be approximated as a continuous medium. The applicability of the sine-Gordon model to \YbNiCuAl\ remains a question. 
The lock-in phenomenon at $q=3/8$ and the first order transition to the ferromagnetic state are also important issues to be studied.

\item
Fourth, we observed an anomalous temperature dependence of the propagation vector. 
The $q$ value changes with temperature almost in proportion to the magnetic order parameter, which we ascribed to the change in the exchange interaction with the development of the ordered moment. 
It is not understood, however, why the $q$ value decreases with decreasing $T$ for $x<0.03$, but it is reversed for $x>0.03$. 

\item
Fifth, the helical ordering of the magnetic octupole moment was detected, which arises simultaneously with the helical ordering of the magnetic dipole moment, reflecting the anisotropic charge density of the crystal field ground state of Yb$^{3+}$ consisting mainly of $|7/2, \pm 1/2 \rangle$. 
%The angle between the moments on the neighboring layers, corresponding to these $q$ values, are $98.2^{\circ}$, $74.0^{\circ}$, $66.4^{\circ}$, and $53.4^{\circ}$, respectively. 

\end{itemize}

\section{Acknowledgements}
We acknowledge valuable comments by H. Ohsumi, M. Suzuki, T. Nagao, R. Shiina, A. Leonov, and J. Kishine. 
This work was supported by JSPS KAKENHI Grant Number 15K05175, 26400333, 16H01073, 15H05885, 17H02912, and 16KK0102.  
The synchrotron radiation experiment was performed under Proposals No. 2015B3711, No. 2016A3761, No. 2016B3762, and No. 2017A3787 at BL22XU of SPring-8. 
We also acknowledge supports by JSPS Core-to-Core Program, A. Advanced Research Networks. 
Magnetization measurement with the MPMS and the electron-probe-microanalysis were performed at N-BARD, Hiroshima University.

\appendix
\section{Formalism of Resonant X-ray Scattering}
We use the scattering-amplitude-operator method to analyze the experimental results.\cite{Lovesey96} 
We consider a $2\times 2$ matrix $\hat{F}$, consisting of four elements of the scattering amplitude for 
$\sigma$-$\sigma'$, $\pi$-$\sigma'$, $\sigma$-$\pi'$, and $\pi$-$\pi'$:
\begin{equation}
\hat{F} = \begin{pmatrix} F_{\sigma\sigma'} & F_{\pi\sigma'} \\ 
F_{\sigma\pi'} & F_{\pi\pi'} \end{pmatrix}\,.
\label{eq:scampG}
\end{equation}
By using the identity matrix $\hat{I}$ and the Pauli matrix $\hat{\mib{\sigma}}$, $\hat{F}$ can generally be expressed as 
\begin{equation}
\hat{F} = \beta \hat{I} + \mib{\alpha} \cdot \hat{\mib{\sigma}}
= \begin{pmatrix} \beta + \alpha_3 & \alpha_1 - i \alpha_2 \\
 \alpha_1 + i \alpha_2 & \beta - \alpha_3 \end{pmatrix} \,.
\end{equation}
Once we know the matrix $\hat{F}$, the scattering cross section $(d\sigma/d\Omega)$ can be calculated by 
\begin{align}
\left( \frac{d\sigma}{d\Omega} \right) &= \text{Tr} \{ \hat{\mu} \hat{F}^{*} \hat{F} \} 
\nonumber \\
&=  \beta^{*}\beta + \mib{\alpha}^{*}\cdot\mib{\alpha} + \beta^{*} (\mib{P} \cdot \mib{\alpha}) + (\mib{P} \cdot \mib{\alpha}^{*})\beta \nonumber \\
& \;\;\;\; + i\mib{P} \cdot (\mib{\alpha}^{*} \times \mib{\alpha}) \,,
\label{eq:CrossSec0}
\end{align}
where $\hat{\mu}=(\hat{I}+\mib{P}\cdot\hat{\mib{\sigma}})/2$ represents the density matrix and the 
Stokes vector $\mib{P}=(P_1, P_2, P_3)$ represents the polarization state of the incident photon. 
$P_1$, $P_2$, and $P_3$ represent the degrees of $+45^{\circ}$ ($P_1=1$) or $-45^{\circ}$ ($P_1=-1$) linear polarization, 
right ($P_2=1$) or left ($P_2=-1$) handed circular polarization, and $\sigma$ ($P_3=1$) or $\pi$ ($P_3=-1$) linear polarization state, respectively. 

Using the elements of (\ref{eq:scampG}), the scattering cross section is expressed as 
\begin{align}
\left( \frac{d\sigma}{d\Omega} \right) &= 
\frac{1}{2} \bigl(\, |F_{\sigma\sigma'}|^2 + |F_{\sigma\pi'}|^2 + |F_{\pi\sigma'}|^2 + |F_{\pi\pi'}|^2 \,\bigr) \nonumber \\
 &\;\;\;\; + P_1 \text{Re} \bigl\{\, F_{\pi\sigma'}^*F_{\sigma\sigma'} + F_{\pi\pi'}^*F_{\sigma\pi'} \,\bigr\} \nonumber \\
 &\;\;\;\; + P_2 \text{Im} \bigl\{\, F_{\pi\sigma'}^*F_{\sigma\sigma'} + F_{\pi\pi'}^*F_{\sigma\pi'} \,\bigr\} 
 \label{eq:CrossSec1} \\
 &\;\;\;\; +  \frac{1}{2} P_3\bigl(\, |F_{\sigma\sigma'}|^2 + |F_{\sigma\pi'}|^2 - |F_{\pi\sigma'}|^2 - |F_{\pi\pi'}|^2 \,\bigr) 
 \,. \nonumber
\end{align}
Therefore, the cross section for the incident beam described by $(P_1, P_2, P_3)$ can generally be written as
\begin{equation}
\left( \frac{d\sigma}{d\Omega} \right) = C_0 + C_1 P_1 + C_2 P_2 + C_3 P_3 \,,
\label{eq:CrossSec2}
\end{equation}
which can be used as a fitting function for the $\Delta\theta_{\text{PR}}$ scan with four parameters of $C_n$ ($n=0\sim 3$). 

The four elements of Eq.~(\ref{eq:scampG}) at an X-ray energy $E=\hbar\omega$ near the resonance energy is expressed as 
\begin{equation}
F_{\varepsilon\varepsilon'}=F_{E1,\varepsilon\varepsilon'}(\omega) + F_{E2,\varepsilon\varepsilon'}(\omega)\,,
\end{equation}
$F_{E1,\varepsilon\varepsilon'}(\omega)$ and  $F_{E2,\varepsilon\varepsilon'}(\omega)$ for a scattering process from the photon state $(\mib{\varepsilon}, \mib{k})$ to $(\mib{\varepsilon}', \mib{k}')$ are expressed as   
\begin{align}
F_{E1,\varepsilon\varepsilon'}(\omega) &= \sum_{\nu=0}^{2} \alpha_{E1}^{(\nu)}(\omega) \sum_{\Gamma} \mib{X}_{E1,\Gamma}^{(\nu)} (\mib{\varepsilon},\mib{\varepsilon}')\cdot \langle  \mib{Z}_{\Gamma}^{(\nu)} \rangle\;, 
\label{eq:GfacE1}\\
F_{E2,\varepsilon\varepsilon'}(\omega) &= \sum_{\nu=0}^{4} \alpha_{E2}^{(\nu)}(\omega) \sum_{\Gamma} \mib{X}_{E2,\Gamma}^{(\nu)} (\mib{\varepsilon}, \mib{\varepsilon}', \mib{\hat{k}}, \mib{\hat{k}}') \cdot \langle \mib{Z}_{\Gamma}^{(\nu)} \rangle \;.
\label{eq:GfacE2}
\end{align}
$\langle \mib{Z}_{\Gamma}^{(\nu)} \rangle$ represents the structure factor of the rank-$\nu$ multipole moment of the irreducible representation $\Gamma$ in the cubic coordinate. 
$\mib{X}_{\Gamma}^{(\nu)}$ is the geometrical factor corresponding to $\langle \mib{Z}_{\Gamma}^{(\nu)} \rangle$. 
For the rank-1 dipole moment with the $T_{1u}$ representation, 
\begin{align}
\mib{X}_{E1,\text{dip}}^{(1)} &= \frac{i}{\sqrt{2}} ( \mib{\varepsilon}' \times \mib{\varepsilon}) \,, \label{eq:XfacE1dip} \\
\mib{X}_{E2,\text{dip}}^{(1)} &= \frac{i}{2\sqrt{10}} \bigl\{(\mib{k}' \cdot \mib{k})(\mib{\varepsilon}' \times \mib{\varepsilon})
+(\mib{\varepsilon}' \cdot \mib{\varepsilon})(\mib{k}' \times \mib{k}) \nonumber \\
& \;\;\; +(\mib{k}' \cdot \mib{\varepsilon})(\mib{\varepsilon}' \times \mib{k})
+(\mib{\varepsilon}' \cdot \mib{k})(\mib{k}' \times \mib{\varepsilon})\bigr\} \,. \label{eq:XfacE2dip}
\end{align}
See Ref.~\citen{Nagao06} for $\mib{X}_{E2}^{(3)}$. 
The scattering amplitude is obtained by taking the scalar product between $\mib{X}_{\Gamma}^{(\nu)}$ and $\langle \mib{Z}_{\Gamma}^{(\nu)} \rangle$. 

The same result is obtained by using the spherical tensor method.\cite{Lovesey05} 
By transforming $\langle \mib{Z}_{\Gamma}^{(\nu)} \rangle$ in the cubic representation to 
$\langle  T_{q}^{(\nu)} \rangle$ in the spherical representation, 
\begin{align}
F_{E1,\varepsilon\varepsilon'}(\omega) &= \sum_{\nu=0}^{2} \alpha_{E1}^{(\nu)}(\omega) \sum_{q=-\nu}^{\nu} (-1)^q X_{E1,-q}^{(\nu)} \langle  T_{q}^{(\nu)} \rangle\;, \\
F_{E2,\varepsilon\varepsilon'}(\omega) &= \sum_{\nu=0}^{4} \alpha_{E2}^{(\nu)}(\omega) \sum_{q=-\nu}^{\nu} (-1)^q X_{E2,-q}^{(\nu)} \langle  T_{q}^{(\nu)} \rangle \;,
\end{align}
where the geometrical factors of $X_{E1,q}^{(\nu)}$ and $X_{E2,q}^{(\nu)}$ are also expressed in the spherical representation as follows. 
\begin{align}
X_{E1,q}^{(\nu)} &= \sum_{p,p'=-1}^1 \varepsilon'_p \varepsilon_{p'} \langle 1 p 1 p' | \nu q \rangle \,, \\
X_{E2,q}^{(\nu)} &= \sum_{p,p'=-2}^{2} h'_{p'} h_{p} \langle 2 p 2 p' | \nu q \rangle \,, \\
h_{m} &=\sum_{p,p'=-1}^{1} \varepsilon_{p} k_{p'} \langle 1 p 1 p' | 2 m \rangle \,, \\
h'_{m} &=\sum_{p,p'=-1}^{1} \varepsilon'_{p} k'_{p'} \langle 1 p 1 p' | 2 m \rangle \,.
\end{align}

It is important that the spectral functions of $\alpha_{E1}^{(\nu)}(\omega)$ and $\alpha_{E2}^{(\nu)}(\omega)$ are rank dependent.\cite{Nagao06,Nagao10}
In the data analysis, each of them may be approximated by a form 
\begin{equation}
\alpha^{(\nu)}(\omega) = \frac{e^{i\phi_{\nu}} }{\hbar\omega - \Delta + i\gamma}\;,
\label{eq:specfcn}
\end{equation}
where $\Delta$, $\gamma$, and $\phi_{\nu}$ are the resonance energy, lifetime broadening effect, and the phase factor, respectively.

\section{Confirmation of the crystal structure}
\subsection{Crystal chirality}
\begin{figure}[b]
\begin{center}
\includegraphics[width=8cm]{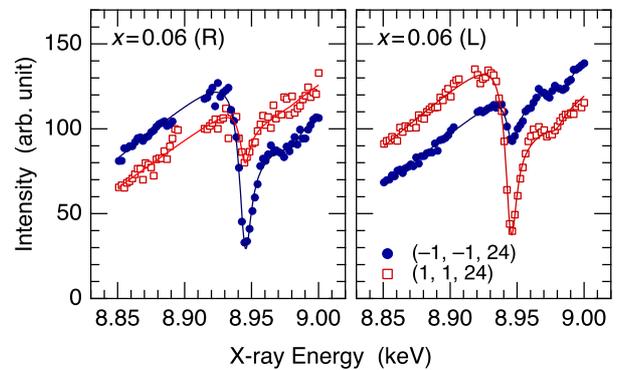}
\caption{X-ray energy dependence of the $(1, 1, 24)$ and $(\bar{1}, \bar{1}, 24)$ fundamental reflections for the right and left handed crystals for $x$=0.06 around the $L_3$ absorption edge of Yb. Solid lines are the guides for the eye. 
}
\label{fig:Edep1124}
\end{center}
\end{figure}

The crystal chirality of the sample, especially the spot where the X ray beam is irradiated in the RXD experiment, can be confirmed at the beam line by measuring the energy dependence of an appropriate fundamental Bragg peak as shown in Fig.~\ref{fig:Edep1124}. 
The structure factor of the $(1, 1, 24)$ fundamental reflection for the right handed crystal is expressed as 
$F_{\text{R}, (1,1,24)}(\omega)=A_{\text{Al}} f_{\text{Al}}(\omega)+A_{\text{Ni}}f_{\text{Ni}}(\omega)+A_{\text{Yb}} f_{\text{Yb}}(\omega)$, 
where $A_{\text{Al}}=0.24+24.6 i$, $A_{\text{Ni}}=-8.6+14.9 i$, and $A_{\text{Yb}}=6.0$. 
$F_{\text{L}, (\bar{1},\bar{1},24)}$ also has the same expression. 
For $F_{\text{L}, (1, 1, 24)}$ and $F_{\text{R}, (\bar{1}, \bar{1}, 24)}$, we take the complex conjugates for $A_{\text{Al}}$ and $A_{\text{Ni}}$. 
When $f_{\text{Yb}}(\omega)=f_{0,\text{Yb}} + f_{\text{Yb}}'(\omega) +  if_{\text{Yb}}''(\omega)$ exhibits an anomalous dispersion around the absorption edge, the intensity of the Bragg reflection also exhibits a strong energy dependence. 
Since the intensities for the right and left handed crystals are expressed as $|F_{\text{R}, (1, 1, 24)}(\omega)|^2$ and $|F_{\text{L}, (1, 1, 24)}(\omega)|^2$, respectively, they exhibit different energy dependences due to the different interference effect.
As shown in Fig.~\ref{fig:Edep1124}, the relative relation of the intensity is reversed for the $(\bar{1}, \bar{1}, 24)$ reflection. 
The relationship of the intensity is consistent the calculated spectrum assuming the predetermined crystal chirality. 

\subsection{Three-fold symmetry about the $c$ axis}
\begin{figure}
\begin{center}
\includegraphics[width=7cm]{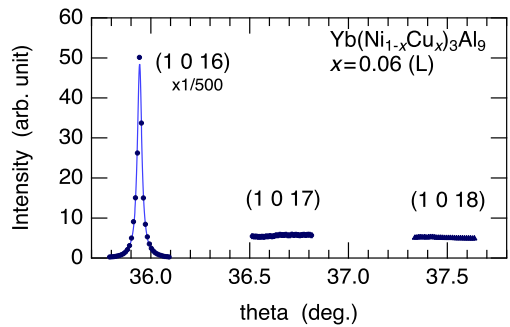}
\caption{Rocking scans of the $(1, 0, 16)$, $(1, 0, 17)$, and $(1, 0, 18)$ reflections for the $x$=0.06 (L) sample. 
}
\label{fig:theta10L}
\end{center}
\end{figure}

Figure \ref{fig:theta10L} shows the rocking scans of the $(1, 0, L)$ reflections for the $x$=0.06 (L) sample. 
Only the $(1, 0,  3n+1)$ reflection is allowed in the  $(1, 0, L)$ reflections in the $R32$ space group. 
These data guarantee the three-fold symmetry of this sample about the $c$ axis. 
If the forbidden reflection is observed, it means that the $[1\, 1\, 0]$ axis is mixed with the $a$ axis due to the stacking fault by $60^{\circ}$. 

\clearpage
%\textbf{Supplemental Material}
\setcounter{section}{19}
\setcounter{figure}{0}

\begin{fullfigure}[b]
\begin{center}
\includegraphics[width=10cm]{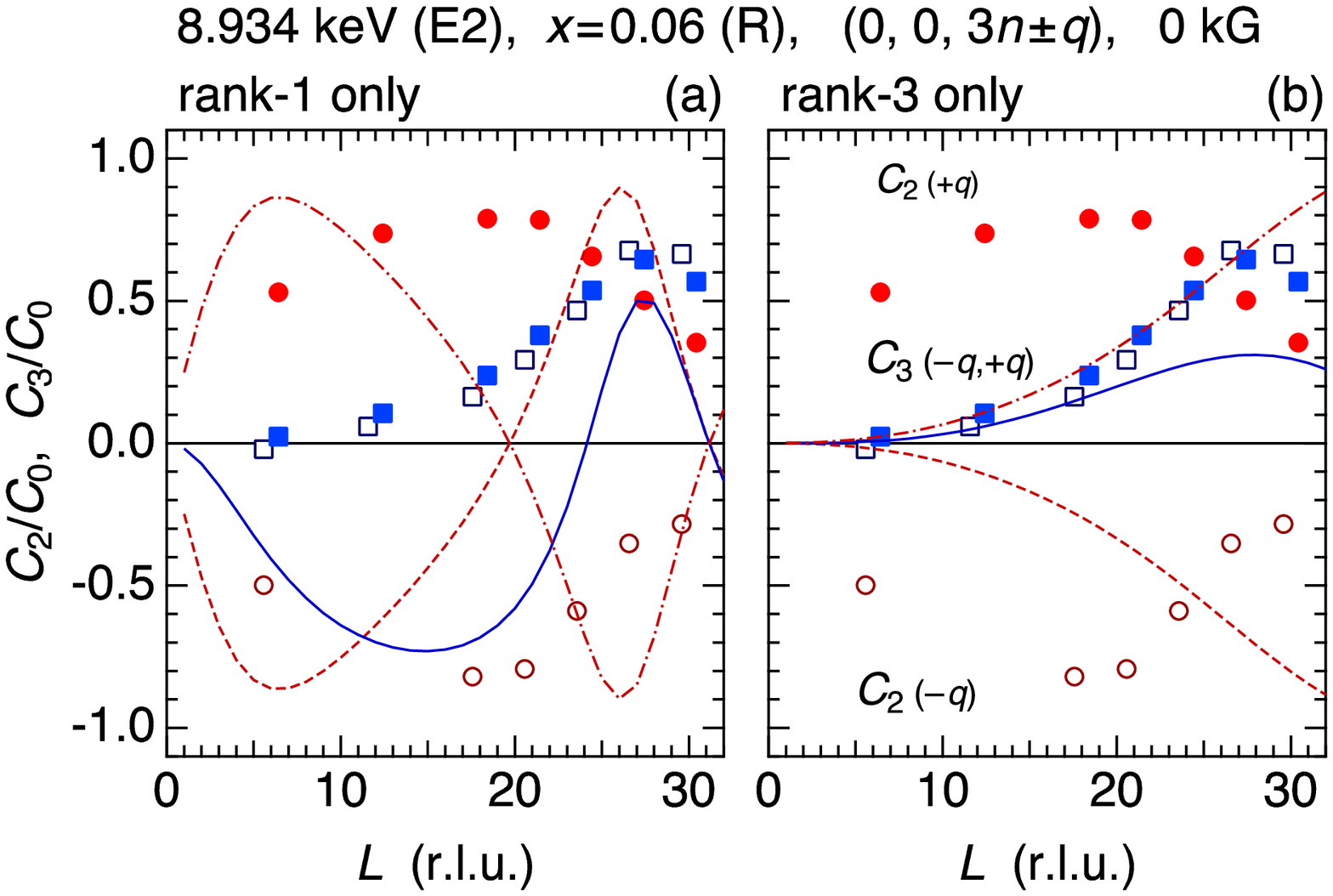}
\caption{Comparison of the disagreeing calculation with the $L$ ($=3n \pm q$) dependence data of the parameters $C_2/C_0$ and $C_3/C_0$ at zero field, which was obtained from the $\Delta\theta_{\text{PR}}$ scans for the $x$=0.06 (R) sample at the $E2$ resonance energy. 
Solid, dashed, and dot-dashed curves represent the calculation of $C_3/C_0$ ($L=3n \pm q$), $C_2/C_0$ ($L=3n - q$), and $C_2/C_0$ ($L=3n + q$), respectively, for models of (a) $\langle J_x \pm i J_y \rangle$ dipole (rank-1) only and (b) 
$\langle T^{\gamma}_x \pm i T^{\gamma}_y \rangle$ octupole (rank-3) only.  The data can be explained by mixing these two scattering contributions as described in the main text.  }
\label{fig:CalcC2C3}
\end{center}
\end{fullfigure}

\begin{fullfigure}[b]
\begin{center}
\includegraphics[width=10cm]{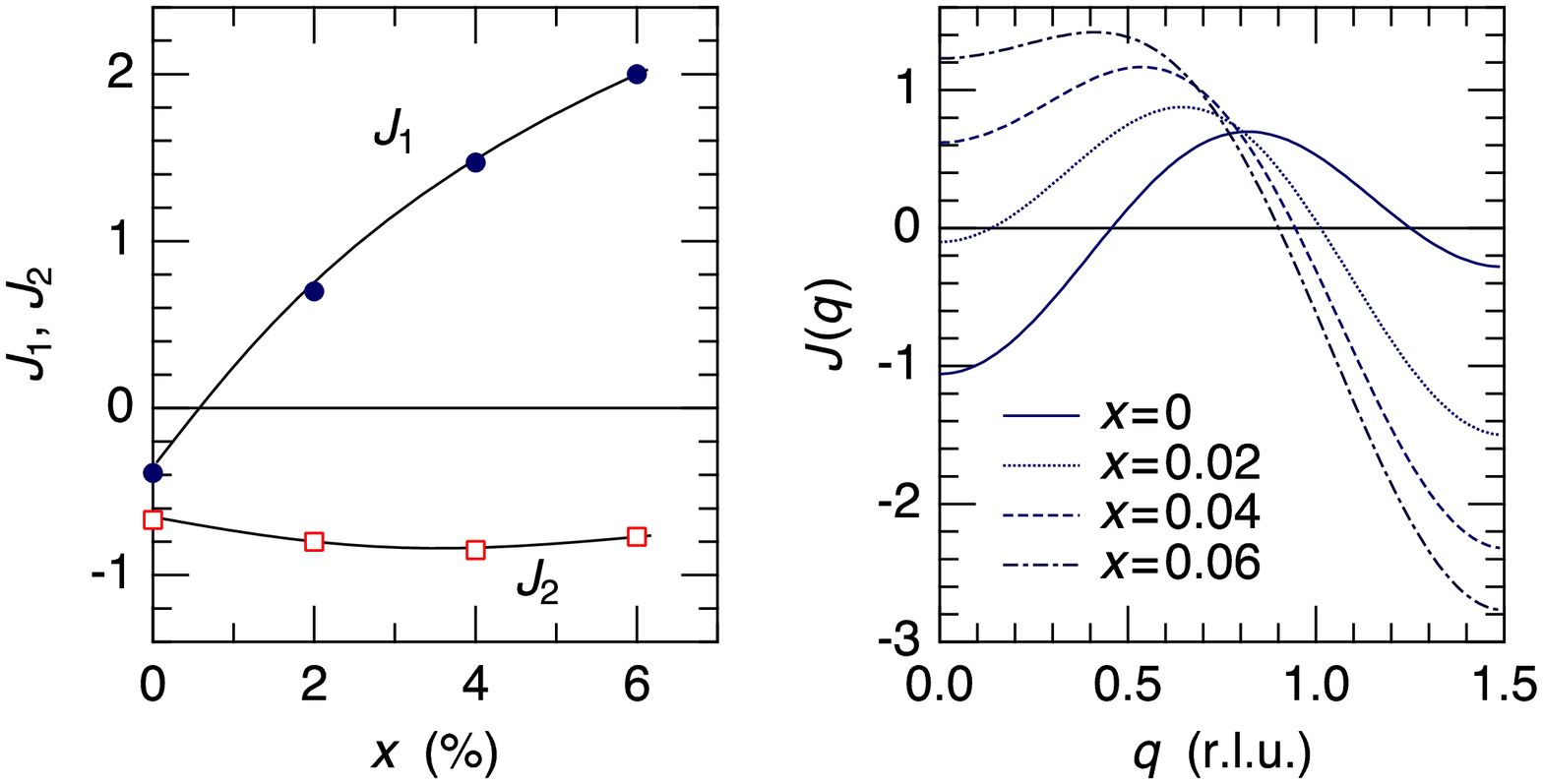}
\caption{A simple calculation of $J(q)$ for one dimensional spin chain with nearest-neighbor exchange $J_1$ and next-nearest-neighbor exchange $J_2$. $J_1$ and $J_2$ are taken so that $q_0=c_0^{-1} \arccos (-J_1/4J_2)$, where $J(q)=J_1 \cos q c_0 + J_2 \cos 2 q c_0$ takes the maximum, reproduces the experimental value and $J(q_0)$ is proportional to $T_{\text{N}}$. The degeneracy at $\pm q_0$ will be lifted by including the DM term. 
In this model, without considering any nonlinear effect such as CSL formation, $H_{\text{C}}$ will be proportional to $J(q_0)-J(0)$. However, it is contradictory to the experimental result. }
\label{fig:J1J2model}
\end{center}
\end{fullfigure}

\begin{table*}[b]
\renewcommand{\arraystretch}{1.5}
\caption{Irreducible representation (Irrep.), operator notation, and basis functions of the dipole (rank-1) and octupole (rank-3) moments in the $O_h$ and $D_3$ point groups. }
\begin{tabular}{cccccccc}
\hline\hline
 & \multicolumn{3}{c}{$O_h$}  & &  \multicolumn{3}{c}{$D_3$} \\
Rank & Irrep. & Notation & Basis Function & & Irrep. & Notation & Basis Function \\
\hline
1 & $T_{1u}$ & $J_x$ & $x$ & & $E_{u}$ & $J_x$ & $x$ \\
 &  & $J_y$ & $y$ & &  & $J_y$ & $y$ \\ \cline{6-8}
 &  & $J_z$ & $z$ & & $A_{2u}$ & $J_z$ & $z$ \\
\cline{1-4}\cline{6-8}
3 & $T_{1u}$ & $T^{\alpha}_x$ & $x(5x^2 - 3 r^2)/2$ & & $E_{u}$ & $T^{\gamma}_x=(-\sqrt{3}T^{\alpha}_x - \sqrt{5}T^{\beta}_x)/2\!\sqrt{2}$ & $\sqrt{6}x(5z^2 - r^2)/4$ \\
 &  & $T^{\alpha}_y$ & $y(5y^2 - 3 r^2)/2$ & &  & $ T^{\gamma}_y=(-\sqrt{3}T^{\alpha}_y +\sqrt{5}T^{\beta}_y)/2\!\sqrt{2}$ & $\sqrt{6}y(5z^2 - r^2)/4$ \\ \cline{6-8}
 &  & $T^{\alpha}_z$ & $z(5z^2 - 3 r^2)/2$ & & $A_{2u}$ & $T^{\alpha}_z$ & $z(5z^2 - 3 r^2)/2$ \\ \cline{2-4}\cline{6-8}
 & $T_{2u}$ & $T^{\beta}_x$ & $\sqrt{15}x(y^2 - z^2)/2$  & & $A_{2u}$ & $T^{\delta}_x=(\sqrt{5}T^{\alpha}_x - \sqrt{3}T^{\beta}_x)/2\!\sqrt{2}$ & $\sqrt{10}x(x^2 - 3y^2)/4$ \\ \cline{6-8}
 &  & $T^{\beta}_y$ & $\sqrt{15}y(z^2 - x^2)/2$ & & $A_{1u}$ & $T^{\delta}_y=(\sqrt{5}T^{\alpha}_y + \sqrt{3}T^{\beta}_y)/2\!\sqrt{2}$ & $\sqrt{10}y(y^2 - 3x^2)/4$ \\ \cline{6-8}
 &  & $T^{\beta}_z$ & $\sqrt{15}z(x^2 - y^2)/2$ & & $E_{u}$ & $T^{\beta}_z$ & $\sqrt{15}z(x^2 - y^2)/2$ \\  \cline{2-4}
 & $A_{2u}$ & $T_{xyz}$ & $\sqrt{15}xyz$ & & & $T_{xyz}$ & $\sqrt{15}xyz$ \\
\hline\hline
\end{tabular}
\end{table*}

\end{document}